\documentstyle[prb,aps]{revtex}

\newcommand{\simless}[0]{\mathbin{\lower 3pt\hbox
   {$\rlap{\raise 5pt\hbox{$\char'074$}}\mathchar"7218$}}} %< or of order

\newcommand{\simgreat}[0]{\mathbin{\lower 3pt\hbox
   {$\rlap{\raise 5pt\hbox{$\char'076$}}\mathchar"7218$}}} %> or of order

\newcommand{\ntotal}[0]{{\cal N}}
\newcommand{\nbulk}[0]{n}
\newcommand{\nsurf}[0]{\theta}
\newcommand{\fracbulk}[0]{g_b}
\newcommand{\fracsurf}[0]{g_s}
\newcommand{\eV}[0]{{\rm eV}}
\newcommand{\cmthree}[0]{{\rm cm}^{-3}}

\begin{document}
\begin{titlepage}
\title{MOLECULE FORMATION IN AND ON GRAINS.I: \\
PHYSICAL REGIMES}
\author{Amir Levinson \cite{Lev}\\ Astronomy Department, Cornell University}
\author{David F. Chernoff\\Astronomy Department, Cornell University}
\author{Edwin E. Salpeter\\Astronomy and Physics Department, 
Cornell University}
\maketitle

\begin{abstract}

We study molecular hydrogen formation in and on solids.  We construct
a model with surface sites and bulk sites capable of describing (1)
the motion and exchange of H and H$_2$ between surface and bulk, (2)
the recombination of H and dissociation of H$_2$ in and on the solid,
and (3) the injection of H from the gas phase and the loss of H and
H$_2$ from the solid.  The basic physical processes include thermally
activated reactions, collisionally induced reactions, and tunneling
reactions.

Our main application is to the astronomical problem of H$_2$ formation
on grains in space but the model has more general applicability.  We
investigate the steady-state H and H$_2$ concentrations in and on the
solid when gas phase atoms or ions stick to the surface or penetrate
the body of the grain.  The model identifies ranges of physical
parameters for which the solid becomes saturated (surface and/or bulk)
with particles (H and/or H$_2$) and facilitates the calculation of the
efficiency of molecule formation (the fraction of the gas phase atoms
that leave as molecules). These solutions are highly degenerate in the
sense that they depend only on a small number of dimensionless
parameters.  We find that a variety of recombination pathways operate
under a broad range of conditions.

As an example we study H$_2$ formation in and on carbon grains. If
molecules form from H on the surface by quantum tunneling {\it alone},
efficient transformation of incoming atoms takes place at grain
temperatures less than $75-100$ K for $1 < (n({\rm H})/\cmthree)
(\epsilon/\eV)^{1/2} < 10^7$ where $n({\rm H})$ and $\epsilon$ are the
number density and energy of incident particles. When incident
particles are energetic enough to penetrate the surface, the bulk of
the grain will be saturated by H and/or H$_2$ under most
circumstances.  Additional molecule formation pathways (recombination
on the surface or in the bulk) increase the efficiency of the
transformation but only rapid bulk to surface exchange of H$_2$ can
alter the conclusion regarding saturation. Saturation can lead to
fundamental changes in the parameters that describe the grain.

\end{abstract}
\end{titlepage}

\section{INTRODUCTION}
\label{sec:intro}

The interaction of hydrogen with solids is a topic of considerable
interest in astrophysics, as well as other applied fields (materials
research in fusion devices, thin film growth and surface etching of
semiconductors, and adsorption and desorption reaction dynamics in
catalysis). There is likewise a great deal of interest in exploring
the fundamental physical processes that govern the interaction between
gas atoms/ions and the solid surface/bulk. In this paper we develop a
model to describe the behavior of H and H$_2$ in and on solids. The
model is phenomenological -- it identifies and parameterizes a set of
reaction pathways thought to play important roles in determining the
concentrations of H and H$_2$ in and on the solid.

Our main application is the study of efficiency of molecule formation
on and release from astronomical grains. The possibility that grain
surfaces catalyze the formation of H$_2$ inside cold atomic/molecular
clouds has long been recognized \cite{HS}.  Under quiescent
conditions, the chemisorption sites are filled and hydrogen is weakly
bound by physical adsorption.  In this scenario, low grain
temperatures are necessary for the catalysis else the H atoms
evaporate before they can recombine. The gas temperature must also be
low else the neutral gas atoms rebound from the grain surface.
Observations show that molecular hydrogen emission is excited
during many phases of a star's life. Well-studied regions include the
Orion Molecular Cloud (Orion OMC-1, e.g. refs. \cite{Ga,Bu},
evolved supernova remnants (IC443 \cite{Tr}; Cygnus
\cite{Gr}), various Herbig-Haro objects (HH 7-11, e.g., refs.
\cite{Ze,Bu}, Seyfert galactic nuclei (NGC1275 \cite{Fi}) and
interacting starburst galaxies (NGC3690-IC694 \cite{Fis}). In each
cited example the emission has been linked to shock heating, however,
shock models often run into a common difficulty, finding a way to
account for the emitting molecular hydrogen under decidedly
non-quiescent conditions.

The main motivation is to explore new and different pathways for H$_2$
formation that may operate under more extreme conditions than hitherto
considered, e.g. in the vicinity of a 100 km s$^{-1}$ shock.  Several
possibilities are of interest: (1) Energetic protons can penetrate the
grain lattice and diffuse within.  H can recombine within the grain or
diffuse to the surface and recombine.  (2) Energetic particles can
drive endothermic molecule formation reactions and also eject and/or
dissociate H containing species from the grain. (3) Energetic
particles can sputter the surface and generate a large surface density
of pits and edges, sites which can act as points of increased physical
adsorption.  We explore the important physical parameters that
regulate molecule formation in the first two scenarios. These
possibilities may turn out to be of astrophysical importance if they
can provide means for rapidly reforming a small amount of molecular
hydrogen in gas that has recently been shocked.  If fast shocks
radiate significantly more molecular line radiation than is currently
thought, re-interpretation of a large body of work may be necessary.

Motivated by these considerations, we explore grain mediated molecule
formation under a wide range of conditions including those that would
be found in the vicinity of a shock.  We describe the most important
physical parameters, we review which parameters have been determined
by laboratory experiments and astronomical observations, and we
construct models that illustrate the different qualitative outcomes
for molecule formation.  Although our analysis is motivated primarily
by astrophysical considerations, it is sufficiently general to be
applied to other systems. For example, modeling high fluence hydrogen
implantation in solids may follow the general strategy presented in
this paper.

In \S \ref{sec:char} we provide a brief overview of the typical energy
barriers of interest in the formation pathways and, for comparison,
the particle energy scales expected behind an astrophysical shock. We
describe the typical grain temperature and note the wide range of
grain composition and size with which we are concerned. In \S
\ref{sec:key} we outline the molecule formation pathways and identify
key parameters. We briefly review the following topics: the
transmission and reflection of incoming gas particles, the stopping
ranges, the possible ordering of energy levels between the bulk,
surface and vacuum, the H-H pairwise interaction in and on the solid,
and the diffusion of H in and on the solid. Where necessary, we adopt
a more or less phenomenological description of rate coefficients for
the succeeding analysis.  In \S \ref{sec:model} we formulate the
model, including the basic equations and the conservation laws. We
focus on a grain with one type of surface site and one type of bulk
site. H$_2$ formation is calculated in three successively more general
models: (i) pairwise recombination involving H$_2$ ejection from the
surface, (ii) plus recombination and binding of H$_2$ on the surface,
(iii) plus recombination and binding in the bulk.  In \S \ref{sec:app}
we apply our results to graphite grains. In \S \ref{sec:sum} we
summarize.

\section{Characteristic Postshock Energy Scales}
\label{sec:char}

Three key endothermic processes -- ion implantation in the solid,
sputtering of the surface atoms and chemical reactions
driven by energetic particles --
occur in the molecule formation pathways.  As we discuss
in more detail below, the threshold
for implantation is expected to be $\sim10$ eV, but has not been
well-characterized for astrophysical grain material. The sputtering
thresholds for H (He) range from $\sim25-60$ eV ($\sim10-16$ eV) for
refractory grain material (graphite, silicate and iron \cite{DS}).  The 
strength of a typical H bond is $1-5$ eV (chemical) or $\sim0.05-0.5$ eV 
(physical absorption, \cite{DuLy}).  The equipartition thermal energy 
of particles behind a 100 km s$^{-1}$ shock is $\sim 10$ eV and 
several factors may increase the
energy of gas-grain collisions. As the grains of size $a$ interact
with the plasma they typically acquire a negative charge $Q \approx
-2.5akT/e$  \cite{Sp} (a number of important additional effects
are noted in ref. \cite{McK}) so that protons strike grains with a total
energy of $\sim35$ eV. In addition, grains of sufficiently large size
($a \simgreat 10^{-5}$ cm) are betatron accelerated behind a strong
radiative shock to velocities $\sim 3 v_s$, where $v_s$ is the shock
velocity  \cite{McK}. From the shock front to the
recombination region, the grains interact with H$^+$, He$^+$ and
He$^{++}$ with typical energies from 1-300 eV. The characteristic
timescale for the gas to cool to $10^4$ K is $t_{cool} \sim 2 \times
10^{10} v_{s7}^{3.2}/n_0$ s  \cite{HMcK}, where $n_0$ is the preshock H-nuclei
density and $v_{s7} = v_s/10^7$ cm s$^{-1}$. 

Observations show that interstellar grains span a range in size at
least as wide as $5 \times 10^{-7} < a < 2 \times 10^{-5}$ cm and have
absorption/emission features typical of silicate (Si-O stretching and
O-Si-O bending modes), graphite and hydrocarbons (C-H stretching
modes) \cite{Dr94}.  

The ratio of the grain temperature to the hydrogen binding energy is
an essential parameter in all molecule formation schemes; it regulates
the rate of thermally activated diffusion, recombination and
evaporation of atoms and molecules from the surface.  The grain
temperature is largely governed by the balance of heating of
Lyman-alpha radiation trapped in the vicinity of the shock front with
cooling by infrared emission.  Silicate grains reach $T_{gr} \approx
320 {\rm K} \left[ (n_0/10^6) v_{s7}^3/ a_{-5} \right]^{0.2}$ and the
results for graphite grains are comparable \cite{HMcK}. Note that at
low densities, $T_{gr}$ is much less, e.g.  $T_{gr} \approx 20 {\rm K}
\left[ n_0 v_{s7}^3/ a_{-5} \right]^{0.2}$ and quantum mechanical
tunneling competes with the mechanism of thermal activation in the
processes listed above.

\section{Key Parameters}
\label{sec:key}

\subsection{Outline for Molecule Formation}

Here we provide a schematic overview of the formation mechanism:

1. H or H$^+$ impinges on a grain; if H$^+$, it is neutralized
by electron transfer as it approaches.

2. A collision takes place and the atom rebounds, sticks or
penetrates the surface. The lattice may be left intact, sputtered or
damaged (i.e. defects introduced). If the grain surface includes
H physically or chemically bound (X-H), then the incoming particle
may drive a ``pickup'' reaction of the form
X-H $+$ H $\to$ X $+$ H$_2$.

3. Within the grain, the atom diffuses from site to site. It seeks out
traps (vacancies, interstitials, etc.) where its binding energy is
greatest. An equilibrium distribution is eventually formed in
which the tightest binding sites are preferentially occupied.

4. When a new atom enters a grain it explores the grain interior
by diffusion. Either it reaches the surface or it finds another atom,
overcomes the activation barrier and forms H$_2$.

5. Bulk H$_2$ diffuses out of the grain, or remains in situ
until it can escape directly. This is possible when damage of
the lattice (by implantation, sputtering, radiation damage or
grain-grain collisions) has accumulated to the point that passageways
to the surface form.

6. If H$_2$ formation in the bulk is energetically unfavorable or inefficient,
then the newly trapped atom diffuses to the surface where it recombines with 
another atom to form an H$_2$ molecule which immediately escapes from the 
grain.  If the grain is too cold for surface recombination to proceed by thermal
activation then quantum tunneling may allow recombination to proceed \cite{Sa}.

7. In the case in which the ambient atoms are not energetic enough to 
penetrate the grain, the H$_2$ formation scenario suitable for cold clouds 
\cite{HS} takes place, but at possibly higher grain
temperature and in the presence of enhanced binding sites (referred to
as ``semi chemisorption'' in that paper). Atoms fill up the tightest
binding sites, until the recombination rate plus the atom's
evaporation rate balances the flux of incoming particles.

The chemical network is governed by a number of key parameters: the
mean interception time of H nuclei by the grain, the diffusion time of
H and H$_2$ inside the grain, the surface and bulk binding energies
for H and H$_2$ and the energy barriers for H$_2$ formation. In
addition to normal sites in and on the grain, there are likely to be
impurity sites whose number and characteristic binding energy for H
and H$_2$ are important.

In the succeeding subsections we review some of the information
relevant to the determination of these parameters and, ultimately, to
our characterization of the rate coefficients for the model. In
section \ref{sec:model}, we examine in detail how the efficiency of
molecular hydrogen formation depends on such quantities.

\subsection{Transmission and Reflection of Low Energy Particles} 
\label{sec:trans}

At low energies (E $\simless 100$ eV) the {\it ab initio} calculation
of the transmission and reflection coefficients of particles from a
target surface is a formidable task.  The results are of importance in
plasma wall experiments, and such calculations have been the subject
of many recent investigations \cite{Ba,EB}.  Many-body effects
play an especially important role. Firstly, in contrast to the
scattering of high-energy particles, which can be well approximated as
a pure binary collision between the projectile and a surface atom, the
scattering of low-energy particles involves simultaneous interaction
with several nearby atoms, owing to the relatively long-range
force between the ion and each
surface atom (e.g., refs. \cite{TSGH}).  As a result, the scattering
of low energy ions is sensitive to the details of the ion-surface
interaction potential.  Secondly, in the case of scattering from metal
surfaces, collective effects produce a significant attractive force,
commonly represented as an image force \cite{TSGH,AC}, which
plays a major role in determining the transmission coefficient at low
energies.

Ignoring chemical reactions, there are three possible outcomes for a
fast particle incident on a surface: the incident particle may stick
to the surface of the target material, penetrate below the surface or
reflect back into space. The typical energy at which each outcome is
maximized varies with the incident ion and the target material.
Masel \cite{Masel} reviews the status of trapping and sticking on 
solid surfaces.  Baskes \cite{Ba} performed simulations of hydrogen 
reflection from a clean nickel surface, using the Embedded Atom 
Method to handle the
many-body interactions in a self-consistent manner.  For normal
incidence, he found that the reflection coefficient peaks at an energy
of order $6$ eV where about $90$\% of the incident ions are
reflected.  Well above $6$ eV the ions are energetic enough to
penetrate the surface, and below it the ions are trapped, owing to the
attractive image force which tends to increase the energy loss of the
scattered ion to the surface.  Similar theoretical results were
obtained by Eckstein \& Biersack \cite{EB} who employed a modified
TRIM code \cite{BE} to calculate hydrogen reflection.  They treated
the surface binding energy as a free parameter and showed that at
energies well above a few eV the reflection coefficient is independent
of the surface potential, but at $\simless$ a few eV, depending on the 
surface binding energy,  
becomes sensitive to it.  In experiments similar behavior has 
been observed for scattering of Na$^+$ from Cu \cite{DiR}.

Less is known about the reflection and sticking of H to non-metals.
The sticking probability of atomic hydrogen on graphite at low
(sub-eV) energies has been inferred from measurements to be at least a
few percent \cite{Be}.  TRIM simulations \cite{EB} of reflection of
low energy H atoms from carbon give a maximum reflection coefficient
smaller than from Ni and W, and a slightly higher energy at the
maximum.  These differences are mainly due to the difference in target
mass.  For a surface binding energy of 1 eV about 50\% of
the H atoms are found to be reflected from the carbon surface at the
peak energy of 4 eV.

Let $T(E,a)$ be the fraction of incident particles that stick to or
penetrate and do not exit a grain of size $a$ at a given
energy E.  (Hence $T \sim 0$ for particles whose energy-dependent stopping
range exceeds the grain size; equivalently, $T(E,a) \sim 0$ for $E >
E_{max}(a)$ when $E_{max}$ is the maximum energy particle stopped by
the grain.) Let $\sigma_g$ be the grain's geometric cross section
for intercepting particles (including the focusing effect of Coulomb
scattering), let $\vec v_g$ be the grain's velocity and let $f(v)$ be
the particle distribution function. Then the mean rate for particle
interception by the grain is
\begin{equation}
t_{in}^{-1}=\int d^3v f(v) \sigma_g |{\vec v} - {\vec v_g}| T(E,a) .
\label{eq:mean-rate}
\end{equation}

\subsection{Stopping Ranges of Low Energy Ions in Solids} 

Energetic particles that penetrate into the grain will experience
energy loss due to nuclear scattering and electronic stopping, i.e.,
energy transfer from the ion to the target nuclei and electrons (for a
detailed account of the theory of ion stopping in solids see ref.
\cite{ZBL}).  In order to be trapped inside the grain, the particle's
stopping range must not exceed the characteristic grain size.  At low
energies, the stopping power is dominated by nuclear scattering
\cite{Ur} (see also refs. \cite{NAL} for reviews on the stopping power
of an electron gas).

The energy transferred in a pure binary collision from an incident
projectile having energy $E_0$ and mass $m$, to a target particle of
mass $M$, is
\begin{equation}
\Delta W=4E_0{\mu\over(1+\mu)^2}\sin^2{\theta\over2},
\label{eq:energy-transfer}
\end{equation}
where $\mu=m/M$ is the mass ratio, and $\theta$ is the center-of-mass
scattering angle. The lab frame scattering angle, $\psi$, satisfies
\begin{equation}
\tan\psi={\sin\theta\over\cos\theta+\mu}.
\end{equation}
For a proton scattering off heavy target nuclei ($\mu<<1$), Eq.
\ref{eq:energy-transfer} shows that the fraction of energy lost by the
intruder in one collision is at most of order $\mu$.  Consequently,
the number of collisions required to slow the incident proton from
$E_0$ to final energy $E_f$ will be $\sim \ln (E_0/E_f)/\mu$, and the
corresponding stopping range, assuming that the ion undergoes a random
walk with fixed step size, varies like $\mu^{-1/2}$.  However, as
already mentioned in \S \ref{sec:trans}, at low energies, the scattering 
of an ion from an array of atoms is considerably more complicated than pure
binary collision because the ion interacts simultaneously with several
nearby atoms.  This tends to increase the effective mass of the
target object, thereby reducing the stopping power.  Moreover,
screening plays an important role at these energies and needs to be
taken into account properly.  Quantitative determination of range
distributions requires numerical simulations \cite{Ur}.  For our
purposes, however, a rough upper limit on the average stopping range
should suffice.  For incident energies $E_0 < 100$ eV, final energy
$E_f \sim 1$ eV, $\mu \sim 1/12$ (e.g. graphite grain), we estimate
$\le 70$ collisions. If the mean free path is not greater than the
lattice spacing then the stopping range of H will not exceed $\sim 8$
atomic layers.  This upper limit is in agreement with recent
experimental results reported in ref. \cite{LR}.

We assume the grains are large enough to stop the particles that
penetrate and we treat the system as a semi-infinite sample.

\subsection{Energy Levels}
\label{subsec:E-Lev}

The ground state energy of H and H$_2$ in the bulk, on the surface and
in the vacuum are the fundamental parameters that govern
the behavior of hydrogen with respect to solids. We denote the ground state energy
of species X in the bulk, on the surface and in the vacuum as E[X]$^i$
where $i = b$, $s$, and $v$ respectively. For future use we define
\begin{itemize}
\item Chemisorption energy (for H) $E_c = E[{\rm H}]^s - {1 \over 2} E[{\rm H}_2]^v$.
\item Solution energy (for H) $E_s = E[{\rm H}]^b - {1 \over 2} E[{\rm H}_2]^v$.
\item Dissociation energy of H$_2$ (per nuclei) in bulk, on surface and in vacuum 
	$E_d^i = E[{\rm H}]^i - {1 \over 2} E[{\rm H}_2]^i$.
\item Embedding energy (for H) $E_e = E[{\rm H}]^v - E[{\rm H}]^s$.
\end{itemize}
The most basic properties of the solid are governed by the signs and
sizes of these energies. We adopt as the zero of the energy scale the
level associated with H$_2$ in the vacuum, i.e. $E[{\rm H}_2]^v = 0$,
so it follows E$_d^v = 2.24$ eV.  For an arbitrary grain there
remain 4 quantities to specify. A simple counting shows that there are
$3 \times 4 \times 5 \times 6 = 360$ possible orderings of the
remaining levels, so it is impractical to consider separately all
relative arrangements. Even for H levels alone, there exist $3 \times
4 = 12$ possible arrangements (relative orderings) of the levels in
the bulk and on the surface.  However, general arguments regarding
H$_2$ formation and release highlight a much smaller group of
relevant, distinct orderings. We discuss some of these below. Note
that our method of computation (section \ref{sec:model}) does not
depend upon an assumed ordering. Further, it should be clear that in
counting the distinct possibilities above, we have ignored many
additional energy scales that can play important roles, such as the
height of barriers for site-to-site migration within the bulk and the
height of barriers for bulk-to-surface and surface-to-vacuum
transitions in the potential energy curves.

\subsubsection{H in bulk and surface; no H$_2$}

We begin by reviewing the chemisorption and solution energy scales for
H. These are of considerable interest in a number of areas, especially
fusion research on plasma-surface interactions.  New
experimental and theoretical techniques have led to progress in
understanding of the physics and quantitative calculation of the
surface interaction \cite{Ruette,Masel}. 
In many cases, the electronic
wave function of the H atoms tends to form a chemical bond with the
surface atoms, leading to a strong attractive force between the H atom
and the solid surface as the atom approachs.  For metals, the
chemisorption sites are typically deeper than bulk sites (with the
possible exception of deep trapping sites due to vacancies or
defects).  For semiconductors, bonds of a more
local nature are formed and, for insulators, the situation is not
well-understood \cite{Masel}. In these cases and for
graphite, the energy of chemisorption sites can lie above the energy
of bulk sites.

Let us consider the possibilities when the H levels satisfy the
inequality $\{E_c,E_s\} < E[{\rm H}]^v$, which is typically the case
for solids.  The ordering implies that the surface and bulk ground
states of H are energetically stable with respect to H in the vacuum.
The inequality drives the accumulation of H in or on the grain and the
resultant high densities may promote H$_2$ formation by recombination
from surface sites.  Of course, H may escape by thermal evaporation,
by collisional ejection or by molecule formation and subsequent
ejection. If H and H$_2$ loss processes are slow then saturation
occurs which will generally lead to changes in the solid's properties
including energy levels. Given our assumptions above, there exist a
total of 6 distinct level orderings.

For metals, it is typically the case that $E_c < E_s$ \cite{Ch} and
the three possible 1D energy diagrams, shown in figure~\ref{fig:1}, have (i) $0
< E_c < E_s$, (ii) $E_c < 0 < E_s$ and (iii) $E_c < E_s < 0$.  The
solid line represents the schematic interaction potential of a
single H atom with the solid.  The figures illustrate the
chemisorption energy $E_c$, solution energy $E_s$, and the vacuum
dissociation energy of H$_2$ (per H) $E_d^v$.  The two dashed lines
represent half the energy of an H$_2$ pair at a
given distance from the surface.  One dashed line applies to a pair at
fixed interatomic separation (equal to the bond length of a free H$_2$
molecule $\sim0.74$ \AA) and the other to the minimum energy
configuration. The minimum energy configuration terminates near the
surface where the pair separation becomes large.  In general, the sign of the
energy is defined by the direction of the corresponding arrow;
positive if the arrow points upwards and negative if it points
downwards.

It is observed that (e.g, \cite{Ch}, and references therein) as an
energetic incoming H$_2$ molecule approaches the solid, it may
dissociate into H atoms that are bound to the surface (with activation
energy barrier per H of $E_a$); in the inverse process, pairs of H
atoms on the surface may tunnel through or thermally ascend the energy
barrier and recombine as H$_2$ (with recombination energy barrier per
H of $E_r$). In figure~\ref{fig:1} vertical bars without arrows represent
manifestly positive quantities.  We note that quantum pair
recombination from the surface is allowed only in case (i).  Surface
recombination requires thermal activation in cases
(ii-iii).\footnote{Tunneling of a hydrogen pair from bulk to vacuum is
energetically possible in cases (i-ii) but not (iii). However, in case
(ii) the decay time to a surface site is likely to be very short and
the corresponding transition rate negligibly small. Differences in
H$_2$ formation between cases (i) and (ii) may be important but are
unlikely to be influenced by the bulk-vacuum tunneling rate.} In the
case of metals, the embedding energy $E_e = E[{\rm H}]^v - E_c$
typically lies in the range 2.4 to 2.8 eV \cite{Ch}, implying $E_c<0$
[case(ii) or (iii)].

The situation for carbon and other non-metals is not as well
characterized. Schermann et al. \cite{Sc} have reported the detection
of H$_2$ formation by recombination of H atoms adsorbed on a carbon
surface at temperatures between 90 and 300 K. The detected molecules
appear to be in highly excited vibrational states. At zero temperature
the measured vibrational states of the newly formed molecules
constrain the embedding energy $E_e <1$ eV for the H binding sites on
the surface; consequently $E_c = E[{\rm H}]^v - E_e > 1.24$ eV.  The system
is so cold that thermal effects cannot significantly change these
estimates although there remains the possibility that H may bind to
more than one type of surface site (for example, one physical and the
other chemical). The solubility data implies $E_s<0$ (see
\cite{Wi} and references therein) so that $E_s < 0 < E_c$, illustrated
as case (iv) in figure~\ref{fig:1}. It is plausible that recombination
proceeds via quantum tunneling in this experiment.  The possibility of
quantum pair recombination may have important implications for H$_2$
formation and release which are considered below.

Two other orderings ($E_s < E_c < 0$ and $0 < E_s < E_c$) are possible
but we are unaware of representative materials for which these cases
would be relevant.

\subsubsection{H and H$_2$ in bulk and surface}

We now review what is known about the H pair interaction on the solid
surface and in the bulk. It could be dramatically different than that
in vacuum as it depends critically on many body interactions.  For most
metals as well as carbon, solubility data for hydrogen at sufficiently
low bulk concentrations appear to follow Sieverts' law
\cite{MR}; that is, the bulk concentration of hydrogen is
proportional to the square root of the equilibrium pressure.  This
implies that H$_2$ is not the preponderant state inside the solid,
rather H (or one of its charged states) is. Some recent theoretical
studies \cite{DF} have shown that the interaction between hydrogen
atoms on metal surfaces and between atoms in surface and subsurface
sites is repulsive and short-range with interaction energies of order
0.1 - 0.4 eV between atoms in nearest-neighbor sites.  These facts
hint that there is no bound state of H$_2$ in the bulk and/or that
there are significant energy barriers that prevent the
association of H to form H$_2$ in the bulk. The fact that Sieverts'
law is satisfied even at very high temperatures in certain materials
\cite{SYM} suggests that energy barriers are not solely responsible.
On the other hand, high fluence hydrogen implantation experiments in
graphite and some amorphous carbon films \cite{MS} suggest that H$_2$
formation may take place in the bulk after becoming saturated.  This
observation is not decisive since other interpretations cannot be
ruled out, particularly in view of the high porosity of these
materials. And, in any case, the pair interaction in the solid is
undoubtably density dependent.  At higher concentrations, the
solid-hydrogen systems tend, in some cases \cite{MR}, to undergo phase
transitions to more ordered phases.  We shall not consider such
complications in the present analysis.

Given the great degree of uncertainty we regard the binding energy of
H$_2$ in the bulk and on the surface as free parameters. For illustration,
let us consider the possibility that there exist bound ground states
of H$_2$ (for a not too large separation) in or on the solid.  We
adopt the inequality $\{E_c, E_s\} < E[{\rm H}]^v$ discussed above for
H.  Furthermore, we consider only cases in which $E_d^b >0$ and
$E[{\rm H}_2]^b > 0$ if H$_2$ levels exist in the bulk and similarly
$E_d^s > 0$ and $E[{\rm H}_2]^s > 0$ if H$_2$ levels exist on the
surface.  These inequalities imply that H$_2$ is energetically stable
against dissociation in situ and that escape to the vacuum
is energetically possible.

If the condition that H$_2$ be bound in situ (with respect to H) is not
satisfied, the existence of metastable bound states
may still be possible in principle.  The lifetime of
such states depends on the characteristics of the corresponding energy
barriers and is accounted for by the analysis presented in 
\S \ref{sec:model}.  And, even if the condition is satisfied dissociation 
will still occur due to thermal activation and collisions with injected 
atoms, as discussed in greater detail in \S \ref{sec:model}.

If the energy level of H$_2$ in the bulk and/or on the surface lies
below that of H$_2$ in the vacuum, we may anticipate that the grain
will become saturated with the molecules at sufficiently low
temperatures.

For each of the previously discussed cases (i-iv) one can identify all
the possibilities consistent with the general assumptions. Only case
(i) with $0 < E_c <E_s$ allows H$_2$ levels {\it both in and on} the grain.
(Case [ii] allows H$_2$ levels in but not on the solid, case [iv] on
but not in, and [iii] neither.) Figure~\ref{fig:2} illustrates one of three
possible orderings for case (i).  The thick arrows label different
energetically allowed pathways for H$_2$ formation and release: H$_2$
formation in the bulk and on the surface, surface recombination, and
evaporation of H$_2$ from the surface.  Each reaction may involve an
activation barrier with some characteristic height and width, as shown
schematically in figure~\ref{fig:2}.  The association pathways (labeled 1 and
2) depend upon the H-H pair potential in the bulk and on the surface.
The pair recombination from the surface (either quantum mechanical or
thermal, labeled 3) is identical to the pathway discussed in the
previous section. The evaporation of H$_2$ from the surface is labeled
4.

We briefly review what is known about the association and dissociation
kinematics of H$_2$ at the surface (i.e. paths 2 and 3 and inverses).
Direct computations of these processes are difficult and involve
multi-dimensional potential energy surfaces. They have been performed
in a limited form for some hydrogen-metal systems \cite{Le,Fo}.
Molecular beam experiments as well as self-consistent many-body
calculations indicate that for most metals there is an activation
barrier for dissociation of width $\sim 0.6$ \AA~ about 2 \AA~above
the surface \cite{HM,Fo}.  For simple and nobel metals the barrier
ranges from about 0.2 eV (e.g., for Na) to more than 1 eV \cite{HA}.
The repulsive interaction between the molecule and the surface is due
to the molecule being closed-shell, and is similar to the repulsion
found for closed-shell atoms such as He as they approach the surface.
However, in the case of an H$_2$ molecule the antibonding state is
shifted downwards and gradually filled as it approaches closer to the
surface, thereby leading to a weakening of the attractive H-H
interaction and the ultimate dissociation of the H$_2$ molecule
\cite{No}.  For transition metals (e.g, Ni, Pd) the activation barrier
is appreciably smaller - about 0.05 to 0.1 eV - and for some systems
(e.g., Ni(110)) the dissociation is non-activated.  The reason for the
small activation barrier observed in transition metals, as explained
by Harris \& Anderson \cite{HA}, is that s electrons of the metal can
occupy unfilled d states which are far more localized, thereby
reducing the Pauli repulsion between the H$_2$ core electrons and
metal electrons.  Self-consistent calculations \cite{No} suggest that 
some metals may exhibit, in addition to the activation barrier of dissociation,
a barrier for H$_2$ adsorption with a comparable height.  Between the            
two barriers H$_2$ molecules can be trapped for a relatively long
time, and this may account for the so-called molecular precursor
state.  For other materials there is no second barrier, but there is a
small potential well just above the barrier due to van der Waals
forces that can give rise to physisorption of H$_2$ molecules at very
low ($<$ 20 K) temperatures \cite{Ch}.  Large surface coverage seems
to give rise to appreciably larger barriers, at least in the case of
transition metals.  For instance, the barrier for dissociation on a
Ni(100) surface increases from 0.1 eV for a clean surface to about 0.6
eV for a surface with a full hydrogen monolayer coverage \cite{Fo}.
Given these complications, we regard the activation barriers as
parameters.

\subsection{Diffusion of Hydrogen in Solids}
Let the activation energies of diffusion of H and H$_2$ be $E_{DH}$
and $E_{DH_2}$.  Most theoretical and experimental studies of H
diffusion in solids have focused on diffusion in metals.  
We briefly review what is
known about diffusion in solids.  At sufficiently high temperatures,
the dependence of the diffusion coefficient on temperature is well
described by the Arrhenius law:

\begin{equation}
D=D_o\exp(-E_{DH}/kT),
\end{equation}
where the preexponential factor, $D_o$, is temperature independent.
In the most rudimentary model of diffusion (for a review of the theory
of diffusion of hydrogen in metals see e.g., ref. \cite{Ke}), the
interstitial atom is supposed to be localized at or about a given
site.  Diffusion occurs via a sequence of thermally activated jumps
from one site to an adjacent one, in a random walk manner. The
activation energy $E_{DH}$ is connected, in this model, to the
height of the potential barrier.  The jump frequency is
expected to be of order of the zero point frequency for harmonic
oscillations of the atom around its equilibrium position, denoted here
by $\nu_o$.  Letting $\ell$ denote the hopping length,
the preexponential factor can be approximated as

\begin{equation}
D_o\simeq 10^{-3}q(\ell/\AA)^2(\nu_o/10^{13}s^{-1})\ \ \ {\rm cm^2\ s^{-1}},
\label{eq:pre-exp-diff}
\end{equation}
where $q$ is a geometrical factor which depends on the lattice type.  

The values of $D_o$ and $E_{DH}$ have been measured for some materials
using various experimental techniques (for a review see ref. \cite{VA}).  
For most metals, as well as for some nonmetals for
which there are experimental data, $D_o$ lies in the range $10^{-4}$
to a few times $10^{-2}$ cm$^2$ s$^{-1}$, in accordance with Eq.
(\ref{eq:pre-exp-diff}).  The activation energy appears to vary
appreciably from one substance to another, ranging from about $0.043$
eV for vanadium \cite{VA} to $\sim$ 0.5 eV for
titanium \cite{Wi}.  It should be noted that most of those
measurements have been taken at relatively high temperatures and in a
fairly narrow range.  Moreover, for some materials (e.g., iron) there
are quite large uncertainties, particularly at low temperatures.  The
determination of hydrogen diffusivity in carbon is difficult because
of the presence of deep trapping sites that appear to control the
mobility at low hydrogen concentrations; the diffusion energy inferred
from experiments using low concentrations of hydrogen is about 4 eV,
which is comparable with the trap energy.  The best estimate of
hydrogen diffusivity in graphite (when trapping sites are excluded) 
according to Causey \cite{Ca} is $E_{DH}\simeq 2.8$ eV, 
$D_o\simeq 0.93$ cm$^2$ s$^{-1}$.

At low temperatures ($\sim$ 200 K for metals) quantum effects
play a role, and deviations from the Arrhenius law are expected,
and in some cases have been observed.  (1) At extremely low temperatures, 
the interstitial eigenstates should fulfill Bloch's theorem and form a
band.  The diffusion process is then band propagation similar to
electron conduction.  The rate, limited by interaction
with phonons and lattice defects, decreases as temperature rises.
No indications of band propagation for hydrogen have
been observed yet \cite{Ke}.  (2) At somewhat higher temperatures, the decay
rate of the band states increases, and ultimately approaches the band
width.  As a consequence, at temperatures larger than
some critical temperature the interstitial will be localized about a
specific site, as discussed above.  In this temperature
regime, overbarrier transitions in the manner described above are
negligible, and thermally activated tunneling from one site to another
may dominate the diffusion process.  In the  
small polaron theory \cite{Ke} one treats the
tunneling matrix element as a small perturbation in the total
Hamiltonian.  It is then possible to compute the transition rate
(i.e., hopping frequency) using time-dependent perturbation theory
\cite{Ke}.  The results of such calculations show that well below the 
Debye temperature the transition rate should satisfy a $T^7$ law,
which has not yet been verified experimentally.  At temperatures greater 
than the Debye temperature, the transition rate obeys the Arrehnius law
with an activation energy that corresponds to the energy difference between 
an occupied site and a vacant site (note that the slope of the log D vs. 
1/T curve is different than that in the high temperature regime where
overbarrier transitions dominate) and a preexponential factor that
involves the tunneling matrix element (and therefore should be
strongly isotope dependent).  

From the foregoing discussion it is expected that at sufficiently low
temperatures the diffusion coefficient will generally have
non-exponential temperature dependence. 
The temperature at which this happens depends on the width of the
potential barrier and other details of the interaction of the H atom
with the lattice.  Unfortunately, measurements of the diffusion
coefficient at low temperatures are extremely difficult, and the value
of $D$ below about 200 K is poorly known.

The mobility of hydrogen on surfaces is some 10 - 15 orders of
magnitudes larger than that in the bulk by virtue of the small
barriers between surface sites.  Recent measurements indicate that
there is indeed a sharp transition for surface diffusion from a high
temperature regime, wherein $D$ obeys the Arrehnius law, to a low
temperature regime, where $D$ is essentially temperature independent.
For example for W(110) at zero coverage this transition occurs at
$\sim150$ K \cite{Au}.  However, there is evidence for a
strong anomalous isotope effect, as well as some other complications,
which are not well understood at present.

At high concentrations, hydrogen diffusion in the solid may be altered
significantly, owing to the increasing strength of the self
interaction between the hydrogen atoms.

In view of the large uncertainty in $D$ referred to above, we shall
treat the mean hopping time between bulk sites, defined as
$t_{h}=\ell^2/D$, as a free parameter.

\section{A Model of Molecular Hydrogen Formation in Grains}
\label{sec:model}
\subsection{Basic Equations}

We distinguish sites according to whether they lie in the bulk or on
the surface of the solid and whether they are elements of the regular
solid (``clean'' sites) or correspond to centers of altered binding
(impurities, vacancies, and defects collectively called
``impurities''). In our model a site may be empty ($\phi$), occupied by a
single H with energy $E[{\rm H}]$ or occupied by H$_2$ with energy
$E[{\rm H}_2]$; we do not allow multiple occupancy or excited energy
levels.  Denote by $\rho_{a\alpha} ({\vec x}, t)$ the number density
of bulk sites of type $a$ ($a$ refers to clean bulk sites or to
impurity sites of type $k$ denoted $I_k$) that are occupied by species
$\alpha$ ($\alpha = {\rm H}, {\rm H}_2$) at position ${\vec x}$ at
time t. Similarly, denote by $\sigma_{a\alpha} ({\vec x}, t)$ the
corresponding areal number density of surface sites.

We assume that the average number of H and/or H$_2$ per grain is much
larger than 1 so that rates of relevant binary reactions can be
meaningfully expressed in terms of the average number densities of the
reactants.

We assume that species diffuse in the bulk and along the surface of
the grain and that exchange occurs
between the gas phase, the solid bulk and the surface.
We express the rate of change of the number densities
by the following coupled set of equations:
\begin{eqnarray}
{\partial \rho_{a\alpha}\over \partial t}-\nabla(D_{a\alpha}\nabla
\rho_{a\alpha})=\dot{\rho}_{a\alpha}, \nonumber \\
{\partial \sigma_{a\alpha}\over \partial t}-\nabla_s(D^s_{a\alpha}\nabla_s
\sigma_{a\alpha})=\dot{\sigma}_{a\alpha},
\label{eq:rate-eqs}
\end{eqnarray}
where $D_{a\alpha}$ and $D^s_{a\alpha}$ are the bulk and surface
diffusion coefficients, respectively, and $\nabla_s$ denotes gradient
along the surface. The terms on the right hand side give the rate of
change of density from external sources (we denote for future use
the injection from the gas phase of species $\alpha$
into sites of type $a$ in the bulk and on the surface by
$s_{a\alpha}$ and ${\hat s}_{a\alpha}$, respectively) and from
internal rearrangements of site occupancies.  Internal changes in the
bulk number densities occur from (i) {\it chemical reactions} taking
place between bulk species or between bulk and surface species, (ii)
{\it particle exchanges} between bulk and surface sites, and (iii)
{\it particle losses} to the vacuum (either via evaporation or
recombination).  Rate coefficients for these processes depend on the
parameters of the grain (temperature, energy levels, and so forth,
including occupation probability).

Let $\delta_{\alpha} = (1,2)$ for $\alpha = ({\rm H}, {\rm H}_2)$.
Summing eqs. (\ref{eq:rate-eqs}) over the entire grain, over all
species and filled sites yields the total rate of change of hydrogen nuclei
\begin{eqnarray}
&{d\over dt}&\left\{\Sigma_{a\alpha}\delta_{\alpha} 
\left(\int \rho_{a\alpha}dV+\int{\sigma_{a\alpha}dS}\right)\right\}
=\nonumber \\
&t_{in}^{-1}& - t_{H-out}^{-1}
              - 2 t_{H_2-out}^{-1}
\label{eq:conserve-eq}
\end{eqnarray}
where $t_{in}^{-1}$ is the rate for atoms to strike the grain (eq.
\ref{eq:mean-rate}), $t_{H-out}^{-1}$ is the loss rate of atoms and
$t_{H_2-out}^{-1}$ is the loss rate of molecules. The loss terms
include loss by evaporation (thermally and quantum mechanically), loss
induced by fast collisions (direct ejections, pickup reactions,
enhanced recombinations) and loss by pair recombination (thermally and
quantum mechanically).  In steady-state the flux of atoms intercepted
by the grain is balanced by the total loss rate of atoms and
molecules.  Efficient molecule formation requires the total loss rate
of molecules by all channels to exceed that of atoms.  The relevant
rates depend on the occupation numbers of the different species, which
are determined by the various physical processes taking place in and
on the grain.  Below we explore the solutions to the above equations
in different regimes of the parameter space, and elucidate the
conditions under which effective molecular formation may take place.

In the examples presented below, we further assume that the occupation numbers
are homogeneous within the bulk and homogeneous on the surface and
that the grain temperature $T_g$ is uniform throughout.
In other words, 
we describe the chemical distribution in terms of two zones (bulk and surface).
This approximation is justified whenever the diffusion timescale is
short compared to all other characteristic timescales (for example,
the injection time) and may also be reasonable under less restrictive
conditions (for example, if the injection process is homogeneous).
We define the total number of grain sites on the surface and in the bulk
(occupied $\alpha$ or empty $\phi$)
\begin{eqnarray}
\ntotal = \sum_{a,\alpha,\phi} \left( \int \rho_{a\alpha} dV +
                             \int \sigma_{a\alpha} dA \right) ,
\end{eqnarray}
the fraction of $\ntotal$ that are in the bulk, of type a, and occupied
by species $\alpha$
\begin{eqnarray}
\nbulk_{a\alpha} = \ntotal^{-1} \int \rho_{a\alpha} dV ,
\end{eqnarray}
the fraction of $\ntotal$ that are on the surface, of type a, and occupied
by species $\alpha$
\begin{eqnarray}
\nsurf_{a\alpha} = \ntotal^{-1} \int \sigma_{a\alpha} dA ,
\end{eqnarray}
and the bulk and surface injection rates
\begin{eqnarray}
S_{a\alpha} = \ntotal^{-1} \int s_{a\alpha} dV \\
{\cal S}_{a\alpha} = \ntotal^{-1} \int {\hat s}_{a\alpha} dA .
\end{eqnarray}
Steady state solutions are obtained by numerically
integrating the time dependent rate
equations (\ref{eq:rate-eqs}) (integrated over the appropriate zone)
until equilibrium is achieved.
For the specific contributions we have
included forward and backward reactions related by detailed balance
for each process that involves thermalized surface and bulk reactants. For
reactions driven by impinging fast particles and for particle loss to
the vacuum we have included only the pathway of interest.  We have not
included any direct bulk to vacuum loss rates even when quantum
tunneling might be possible.

Our analysis becomes inapplicable when the mean residence time of an
hydrogen atom in the grain, $\Delta t$ becomes much shorter than
$t_{in}$.  In this case the probability of finding two atoms
simultaneously inside the grain is approximately $\Delta t/t_{in}$,
assuming that injection of atoms is a Poisson process.  Consequently,
even if the rate of molecular formation is shorter than $\Delta t$, at
most a fraction $\Delta t/t_{in}$ of injected atoms will be converted
into molecules.

\subsection{H$_2$ Formation in a Clean Grain}
We consider the simplest case, namely molecular formation in a grain
free of impurities ($\nbulk_{I\alpha}=\nsurf_{I\alpha}=0$). We begin
by giving explicit expressions for the rate coefficients in terms of
the various parameters involved.

\subsubsection{Reaction Rates}
Below we list the processes included in the model. As a general
strategy, we have parameterized the rates in terms of explicit
functions of $\nbulk_H$, $\nbulk_{H_2}$, $\nsurf_H$ and $\nsurf_{H_2}$
(``fractional occupancies'') times characteristic rate coefficients
(of the forms $t^{-1}$, $\alpha S$).  These latter depend primarily on
$T_g$ and various grain energy scales but also on occupancies when
saturation is approached. (Later results are explicit and most useful
when this implicit dependence on fractional occupancy is absent or
weak but remain correct implicit solutions in all circumstances.)

Let $\fracsurf$ and $\fracbulk$ be the fractions of all sites that are
surface and bulk sites respectively (i.e. $\sim \ntotal^{-1/3}$ and $1
- \ntotal^{-1/3}$, up to a geometrical factor; $\fracsurf + \fracbulk
= 1$). The fractional occupancies satisfy $0 \le \nbulk_H +
\nbulk_{H_2} \le \fracbulk$ and $0 \le \nsurf_H + \nsurf_{H_2} \le
\fracsurf$.  We denote an empty bulk site $b\phi$; the
probability that a bulk site is empty is $B^b=1-(\nbulk_H +
\nbulk_{H_2})/\fracbulk$. Likewise, we denote an empty surface site
$s\phi$; the probability that a surface site is empty is
$B^s=1-(\nsurf_H + \nsurf_{H_2})/\fracsurf$.

The reactions are listed and the forms for the characteristic rate
coefficient are given immediately to the right, followed by a short
description of each process and the expression for the rate of change
(always expressed as per site for the $\ntotal$ grain sites).

\begin{itemize}

\item Particle Exchange

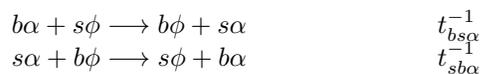
\begin{itemize}

\item
\[ \begin{array}{l@{\hspace{1in}}l}
	b\alpha + s\phi \longrightarrow b\phi + s\alpha &
\mbox{$t_{bs\alpha}^{-1}$} \\ 	s\alpha + b\phi \longrightarrow s\phi
+ b\alpha & \mbox{$t_{sb\alpha}^{-1}$} \end{array} \] Thermally
activated diffusion of a species $\alpha=$ H or H$_2$ from the bulk to
the surface and vice versa.  The fluxes are proportional to
the product of bulk (surface) concentration and the number of vacant
neighbor surface (bulk) sites.  The rates per site may be expressed as
$t_{bs\alpha}^{-1} \nbulk_{\alpha}(\fracsurf B^s/\fracbulk)$, and
$t_{sb\alpha}^{-1}\nsurf_{\alpha}B^b$ where $t_{bs\alpha}$ and
$t_{sb\alpha}$ are characteristic timescales to hop to nearest
neighbor sites. The hopping timescales depend on the grain
temperature, the characteristic binding energies and the form of the
interaction. We assume exchanges between
neighboring lattice sites; order unity range variations are accommodated in the
definition of the characteristic timescale.  Detailed balance implies
$t_{sb\alpha}^{-1}=\exp\{(E^s[\alpha]-E^b[\alpha])/k
T_g\}t_{bs\alpha}^{-1}$.  If $E^b[\alpha]-E^s[\alpha]=0$ then
$t_{bs\alpha}=t_{sb\alpha} \sim t_{h\alpha}$.

\item 
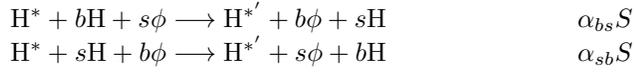
\[ \begin{array}{l@{\hspace{1in}}l}
   {\rm H}^* + b{\rm H} + s\phi \longrightarrow {\rm H}^{*'} 
+ b\phi + s{\rm H} & \mbox{$\alpha_{bs}S$} \\
   {\rm H}^* + s{\rm H} + b\phi \longrightarrow {\rm H}^{*'} 
+ s\phi + b{\rm H} & \mbox{$\alpha_{sb}S$}
   \end{array}
\]
Collisional displacement of H atoms from the bulk to the surface by
energetic incident atoms (H$^*$) and vice versa.  The rate coefficients are 
taken to be proportional to the total injection rate per site, 
$S=1/\ntotal t_{in}$, with efficiencies which are denoted by 
$\alpha_{bs}$ and $\alpha_{sb}$, respectively.  The rate per site 
for transport from bulk to surface takes the form $\alpha_{bs}
S\nbulk_H(\fracsurf B^s/\fracbulk)$ and from surface to bulk, $\alpha_{sb}S
\nsurf_HB^b$. 

The efficiencies depend only on the energy of the incident particle
and the energy level of the various sites as long as the density is
smaller than the inverse volume sampled by the incident particle.
Once the grain becomes saturated the efficiency of collisional
displacement will depend on density, for example, when one dislodged
atom can dislodge additional atoms.

A plausible limiting efficiency may be derived as follows.  If in
slowing down, an incident atom visited every site in the grain once,
then each occupying atom suffers $\sim 1$ displacement. A surface atom
is moved into the bulk with order unity probability; a bulk atom is
moved into the surface with probability $\fracsurf$. From the form of
the above rate expressions, this corresponds to $\alpha_{bs}$ and 
$\alpha_{sb} \sim \ntotal$.  Thus, we assume that the efficiencies 
range from zero to roughly $\ntotal$.

\end{itemize}

\item H$_2$ formation and destruction

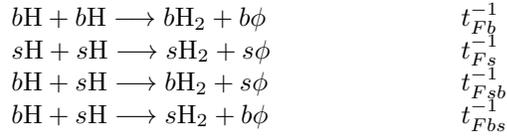
\begin{itemize}

\item 
\[ \begin{array}{l@{\hspace{1in}}l}
	b{\rm H} + b{\rm H} \longrightarrow b{\rm H}_2 
+ b\phi & \mbox{$t_{Fb}^{-1}$} \\
	s{\rm H} + s{\rm H} \longrightarrow s{\rm H}_2 
+ s\phi & \mbox{$t_{Fs}^{-1}$} \\
	b{\rm H} + s{\rm H} \longrightarrow b{\rm H}_2 
+ s\phi & \mbox{$t_{Fsb}^{-1}$} \\
	b{\rm H} + s{\rm H} \longrightarrow s{\rm H}_2 
+ b\phi & \mbox{$t_{Fbs}^{-1}$} 
   \end{array}
\]
Molecular hydrogen formation in the bulk and on the surface (reaction
pathways labeled 1 and 2 in figure~\ref{fig:2}).  We define $t_{Fb}$ ($t_{Fs}$)
to be the H$_2$ formation time, given that two H atoms are in
neighboring, bulk (surface) sites. We define $t_{Fbs}$ ($t_{Fsb}$) to
be the time of H$_2$ formation on the surface (bulk) by recombination
of subsurface and surface atoms.  If H$_2$ formation in or on the
grain is forbidden then $t_{Fi}=t_{Fij}=\infty$, while in the absence
of an activation barrier $t_{Fb}\sim t_{h}$.  Generally $t_{Fi}$ are
finite and depend on the height and width of the corresponding
activation barriers for H$_2$ formation in the solid, as depicted in
figure~\ref{fig:2}.  The rate per site for H$_2$ formation in the bulk is
$[t^{-1}_{Fb}\nbulk_H+t^{-1}_{Fsb}\nsurf_H]\nbulk_H/\fracbulk$,
and that for H$_2$ formation on the surface
$[t^{-1}_{Fs}(\nsurf_H/\fracsurf)+t^{-1}_{Fbs}(\nbulk_H/\fracbulk)]\nsurf_H$.
As above, we have assumed that the hydrogen pair interaction range is one
lattice spacing but the variation may be absorbed into the
definition of $t_{Fi}$.

\item 
\[ \begin{array}{l@{\hspace{1in}}l}
  {\rm H}^* + b{\rm H} \longrightarrow b{\rm H}_2 & \mbox{$\alpha_{Fb}S$}
   \end{array}
\]
H$_2$ formation in the bulk by direct recombination of 
injected atoms with interstitial atoms. The rate of change of 
$\nbulk_{H_2}$ is given by $\alpha_{Fb}S\nbulk_H$ where the efficiency 
$\alpha_{Fb}$ depends only on the energy of the incident atom and the 
site binding energy.  Again, $\alpha_{Fb}$ ranges between zero and 
$\ntotal$.  If H$_2$ formation in the bulk is forbidden, then 
$\alpha_{Fb}=0$.

\item 
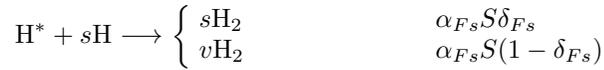
\[ {\rm H}^* + s{\rm H} \longrightarrow
	\left\{ 
		\begin{array}{l@{\hspace{1in}}l}
		s{\rm H}_2 & \alpha_{Fs} S \delta_{Fs} \\
		v{\rm H}_2 & \alpha_{Fs} S (1-\delta_{Fs})
		\end{array}
	\right.
\]
H$_2$ formation on the surface by direct recombination with
the incident atom.  The corresponding rate of change of 
$\nsurf_{H_2}$ is given by $\alpha_{Fs}S\nsurf_H$ where the 
efficiency is $\alpha_{Fs}$.  A fraction $\delta_{Fs}$ of the 
formed molecules are assumed to remain on the surface, while 
the rest are immediately ejected from the grain.  Following 
Duley \cite{Du}, we shall refer to the latter process as prompt 
reaction.  

\item 
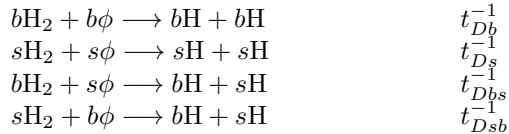
\[ \begin{array}{l@{\hspace{1in}}l}
	b{\rm H}_2 + b\phi \longrightarrow b{\rm H} 
+ b{\rm H} & \mbox{$t_{Db}^{-1}$} \\
	s{\rm H}_2 + s\phi \longrightarrow s{\rm H} 
+ s{\rm H} & \mbox{$t_{Ds}^{-1}$} \\
	b{\rm H}_2 + s\phi \longrightarrow b{\rm H} 
+ s{\rm H} & \mbox{$t_{Dbs}^{-1}$} \\
	s{\rm H}_2 + b\phi \longrightarrow b{\rm H} 
+ s{\rm H} & \mbox{$t_{Dsb}^{-1}$} 
   \end{array}
\]
H$_2$ dissociation in or on the solid owing to thermal activation and
quantum tunneling (in the case where metastable states exist).
Detailed balance implies that the ratio of formation to dissociation
rate $t_{Fi}^{-1}/t_{Di}^{-1} = \exp(2E_d^i/kT_g)$, $i=b,s$.  We 
take the dissociation rate per site to be $t_{Db}^{-1}\nbulk_{H_2}B^b$ 
in the bulk, $t_{Ds}^{-1}\nsurf_{H_2}B^s$ on the
surface. Likewise the rate coefficient for dissociation of H$_2$ in
the bulk giving
surface and subsurface H is
$t_{Dbs}^{-1} =
t_{Fsb}^{-1}\exp\{-(2E^b_{d}+E[H]^s-E[H]^b)/kT_g\}$; for H$_2$ on the
surface the corresponding rate coefficient is
$t_{Dsb}^{-1} =
t_{Fbs}^{-1}\exp\{-(2E^s_{d}+E[H]^b-E[H]^s)/kT_g\}$.
The rates become
$t_{Dbs}^{-1}\nbulk_{H_2}(\fracsurf B^s/\fracbulk)$ and
$t_{Dsb}^{-1}\nsurf_{H_2}B^b$, respectively.

\item 
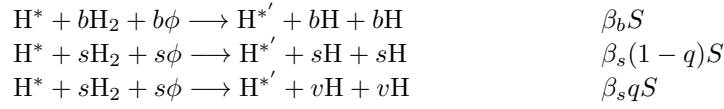
\[
  \begin{array}{l@{\hspace{1in}}l}
	{\rm H}^* + b{\rm H}_2 + b\phi \longrightarrow {\rm H}^{*'} 
+ b{\rm H} + b{\rm H} & \beta_b S \\
	{\rm H}^* + s{\rm H}_2 + s\phi \longrightarrow {\rm H}^{*'} 
+ s{\rm H} + s{\rm H} & \beta_s (1-q) S \\
	{\rm H}^* + s{\rm H}_2 + s\phi \longrightarrow {\rm H}^{*'} 
+ v{\rm H} + v{\rm H} & \beta_s q S
  \end{array}
\]
H$_2$ dissociation in or on the solid by injected atoms.
The bulk and surface dissociation rates per site are 
$\beta_b S\nbulk_{H_2}B^b$ and $\beta_s
S\nsurf_{H_2}B^s$ where $\beta_i$ gives the average number of dissociations
per target molecule per site as the atom slows down for $i=b,s$.  Further, we
suppose that a fraction $q$ of H$_2$ dissociations on the surface results
in the immediate loss of the H atoms thereby produced from the grain.

\end{itemize}

\item Particle Losses

\begin{itemize}

\item Thermal evaporation of H atoms from the grain's surface at a 
rate $\gamma_H\nsurf_H$.

\item Collisional ejection of H atoms from the grain's surface at a 
rate $\alpha_{svH}S
\nsurf_H$.

\item Evaporation of H$_2$ molecules from the surface 
(thermal and quantum traversal of pathway 4 in figure~\ref{fig:2}) at
a rate given by $\gamma_{H_2}\nsurf_{H_2}$.

\item Collisional ejection of H$_2$ molecules from the grain's surface 
at a rate $\alpha_{svH_2}S\nsurf_{H_2}$.

\item Pair recombination of adjacent surface atoms (thermal and
quantum association via pathway 3 in figure~\ref{fig:2}), with a
rate given by $\fracsurf^{-1}\Gamma_{2H}\nsurf_H^2$.  We henceforth refer 
to this process as ``pair evaporation.''

\end{itemize}

\end{itemize}

For our models we need only the {\it sum} of various individual
processes. These are:

\begin{itemize}
\item The rate coefficient for bulk to surface exchange is
$\tau^{-1}_{bsH}=t^{-1}_{bsH}+\alpha_{bs}S$; the total rate
is $\nbulk_H(\fracsurf B^s/\fracbulk) \tau^{-1}_{bsH}$.
Likewise, the rate coefficient for surface to bulk exchange is
$\tau^{-1}_{sbH}=t^{-1}_{sbH}+\alpha_{sb}S$; the total rate is
$\nsurf_H B^b \tau^{-1}_{sbH}$. The rate coefficients for
H$_2$ exchange are completely analogous.

\item The dissociation rate coefficient on the surface is
$\tau_{Ds}^{-1} = t_{Ds}^{-1} + \beta_s S$; in the bulk $\tau_{Db}^{-1}= 
t_{Db}^{-1} + \beta_b S$; and for surface plus subsurface products
$t_{Dbs}^{-1}$ and $t_{Dsb}^{-1}$.  We also define
the part of the surface dissociation rate that leads to
an increase of $\nsurf_H$ (as opposed to immediate escape)
$\tilde{\tau}_{Ds}^{-1}= \tau_{Ds}^{-1}-q\beta_sS $.

\item The rate coefficient for thermal and collisional 
loss of H from the surface (all linear in $\nsurf_H$) is $\Gamma_{H}
=\gamma_{H}+\alpha_{svH}S$. The rate coefficient for H$_2$ loss by
thermal, quantum and collisional loss is 
$\Gamma_{H_2}=\gamma_{H_2}+\alpha_{svH_2}S$.

\item The H$_2$ formation rate in the bulk is
$\nbulk_H (t_{Fb}^{-1} \nbulk_H/\fracbulk 
+ t_{Fsb}^{-1} \nsurf_H/\fracbulk + \alpha_{Fb} S)$; the
thermally driven H$_2$ formation rate on the surface is
$\nsurf_H (t_{Fs}^{-1} \nsurf_H/\fracsurf + t_{Fbs}^{-1} \nbulk_H/\fracbulk)$; 
the H$_2$ formation rate by collisions and by pair evaporation is $\nsurf_H
(\alpha_{Fs} S + \fracsurf^{-1}\Gamma_{2H} \nsurf_H)$.
\end{itemize}

\subsubsection{Relation of Rate Coefficients to Energy Scales}

We now turn to the evaporation rates.  We suppose that the
interaction potential of a single hydrogen atom with the surface is
similar to that shown schematically in figure~\ref{fig:1}, with $E_e>0$.
The evaporation rate of an H atom from a surface site
is

\begin{equation}
\gamma_{H}=\nu_o\exp\{-E_e/kT_g\},
\end{equation}
where $\nu_o\sim10^{13}$ s$^{-1}$ is the characteristic frequency of oscillation
of an H atom about its equilibrium position.

The pair evaporation rate, $\Gamma_{2H}$, depends on the shape of the
pair surface potential.  Atom pairs can escape from the surface by
either hopping above the finite activation barrier or if $E_c > 0$ by
tunneling through it \cite{Sa}.  When $E_c > 0$ recombination can take place
even at zero temperature and tunneling dominates the overbarrier rate at
sufficiently low temperatures. In the following, we assume that the
initial separation between the recombining atoms is of the order of
the H-H interaction range (about 1 - 2 surface sites).

We consider first a grain with $E_c > 0$ at zero temperature, as shown
in figure~\ref{fig:1}(i) and (iv).  Let ${\bf R}$ be the vector position of the
center of mass of the pair system, ${\bf r}$ one-half the relative
position vector of the nuclei, and $V({\bf r},{\bf R})$ the
corresponding adiabatic potential.  To zeroth order, the penetration
factor of an atom pair of energy $E$ along a given path joining the
points $a$ and $b$ in this six-dimensional space, is given in the WKB
approximation by
\begin{equation}
G_{S}=2(2m_{2H}/\hbar^2)^{1/2}\int_{a}^{b}{dS[V({\bf r}(S),{\bf R}(S))-E]^{1/2}},
\end{equation}
where $dS$ is a line element, and $V(a)=V(b)=E$.  The transmission
coefficient can be determined by summing $e^{-G_S}$ over all paths
joining the points $a$ and $b$, and the corresponding tunneling rate
is obtained by multiplying the result by $\nu_o$, the attempt
frequency.  Lagrange's equations describe the path of minimum
penetration factor for fixed endpoints $a$ and $b$.  The dominant
contribution to the tunneling rate comes from that path and to a good
approximation at zero temperature the surface recombination rate is
\begin{equation}
\Gamma_{2H}\simeq \nu_oe^{-G},
\label{eq:zero temp gamma}
\end{equation}
where $G$ denotes the appropriate value of $G_S$. To simplify the analysis
further, we consider only paths along which ${\bf r}$ is constrained
to be parallel to the surface.  We assume that the potential energy
surfaces are co-planer for fixed separation so that the potential
depends only on the two coordinates $z$ and $r$, where $z$ is the
height above the surface (i.e., the component of ${\bf R}$ normal to
the surface) and $r=|{\bf r}|$ is one-half the distance between the
hydrogen nuclei. Then the penetration problem reduces to tunneling
through an effective one-dimensional potential
$v(z)=[V(z)-E][1+(dr/dz)^2]$ where $(dr/dz)$ describes the appropriate
path.  In view of the uncertainties in the 2D potential energy
surfaces, we approximate $v(z)$ by an inverted harmonic oscillator
potential with barrier height $\Delta E$ and width $z_o$. Then
\begin{equation}
G_H=48.5 (\Delta E/eV)^{1/2} (z_o/\AA)
\label{eq:pent. fac.}
\end{equation}
and eq. (\ref{eq:zero temp gamma}) gives the zero temperature
recombination rate.  The result is easily generalized to finite
temperatures.  Averaging the tunneling rate over the Boltzmann
distribution while regarding the transmission coefficient as unity
for $E>\Delta E$ yields
\begin{equation}
\Gamma_{2H}={\nu_o\over G_H-\Delta E/kT_g}\left[G_He^{-\Delta E/kT_g}
-(\Delta E/kT_g)e^{-G_H}\right].
\label{eq:pair-rec}
\end{equation}
The rate is strongly enhanced when the temperature approaches the
tunneling temperature, $ T_t=\Delta E/ kG_H\simeq 2.4\times10^2(\Delta
E/eV)^{1/2}(z_o/\AA)^{-1}$ K, and approaches the classical limit above
$T_t$. 

For materials with $E_c>0$, the pair evaporation rate is given by eq.
(\ref{eq:pair-rec}) with $\Delta E=E_r$ (see figure~\ref{fig:1}[i,iv]), and $G_H$
the penetration factor at energy $E_c$, i.e. the ground state energy
of the trapped particles. The limiting behavior is
\begin{equation}
\begin{array}{ll}
\Gamma_{2H} = \nu_o e^{-G_H \min \left( {T_t \over T_g}, 1 \right) } 
& \mbox{if $E_c>0$.}
\end{array}
\label{eq:pair-rec-limit1}
\end{equation}

For materials $E_c<0$, thermal activation is required (see figure
\ref{fig:1}[ii,iii]).  Only atom pairs with total energy $\ge E[H_2]^v$ can
tunnel through the barrier and we find that the surface recombination
rate is eq. (\ref{eq:pair-rec}) reduced by a factor $\exp(-E_c/kT_g)$.
Here, $\Delta E = E_a$ (see figure~\ref{fig:1}[ii,iii]) and the interpretation of
$G_H$ is that it is the penetration factor of an atom pair having
energy $E[H_2]^v$.  The limiting behavior is
\begin{equation}
\begin{array}{ll}
\Gamma_{2H} = \nu_o e^{E_c \over kT_g} e^{-G_H \min \left( {T_t \over T_g}, 1 \right) } 
& \mbox{if $E_c <0$.}
\end{array}
\label{eq:pair-rec-limit2}
\end{equation}

In the case in which H$_2$ formation in or on the solid takes place,
the evaporation of H$_2$ molecules from the grain surface proceeds in
an analogous fashion. If $E[H_2]^s > E[H_2]^v$, as assumed for figure~\ref{fig:2},
then H$_2$ escapes by both overbarrier and underbarrier transitions.
On the other hand, if $E[H_2]^s < E[H_2]^v$ then H$_2$ requires
thermal activation to escape.  Like $\Gamma_{2H}$, the H$_2$
evaporation rate, $\gamma_{H_2}$, depends on the energy barrier and
whether or not tunneling can occur at zero temperature.

\subsubsection{Rate Equations}

We suppose that the occupation numbers are homogeneous in the bulk and 
(separately) on the surface.  As already mentioned above, this assumption 
is justified when the diffusion time is short enough.  The equations governing 
$\nbulk_H$, $\nbulk_{H_2}$, $\nsurf_{H_2}$ and $\nsurf_H$ are given respectively by,

\begin{eqnarray}
\dot{\nbulk}_H&=&S_H-[\tau_{bsH}^{-1}(\fracsurf B^s/\fracbulk)+2\alpha_{Fb}S]\nbulk_H-2\fracbulk^{-1}t_{Fb}^{-1}\nbulk_H^2
+\tau_{sbH}^{-1}B^b\nsurf_H\nonumber \\
&+&(2\tau^{-1}_{Db}B^b+t^{-1}_{Dbs}\fracsurf B^s/\fracbulk)\nbulk_{H_2}+t^{-1}_{Dsb}
B^b\nsurf_{H_2}-\fracbulk^{-1}(t_{Fsb}^{-1}+t_{Fbs}^{-1})\nbulk_H\nsurf_H,
\label{eq:rate NH}
\end{eqnarray}

\begin{eqnarray}
\dot{\nbulk}_{H_2}=-[\tau^{-1}_{Db}B^b+(\fracsurf B^s/\fracbulk)t^{-1}_{Dbs}&+&
\tau_{bsH_2}^{-1}(\fracsurf B^s/\fracbulk)]\nbulk_{H_2}+\alpha_{Fb}S\nbulk_H\nonumber \\
&+&t^{-1}_{Fb}\fracbulk^{-1}\nbulk_H^2+\tau_{sbH_2}^{-1}B^b\nsurf_{H_2}+
\fracbulk^{-1}t_{Fsb}^{-1}\nbulk_H\nsurf_H,
\label{eq:rate NH2}
\end{eqnarray}

\begin{eqnarray}
\dot{\nsurf}_{H_2}=\tau_{bsH_2}^{-1}\nbulk_{H_2}(\fracsurf B^s/\fracbulk)
&+&t^{-1}_{Fs}\fracsurf^{-1}\nsurf_H^2
+\alpha_{Fs}\delta_{Fs}S\nsurf_H+\fracbulk^{-1}t_{Fbs}^{-1}\nbulk_H
\nsurf_H\nonumber \\ &-&(\tau_{sbH_2}^{-1}B^b
+\Gamma_{H_2}+\tau^{-1}_{Ds}B^s+t^{-1}_{Dsb}B^b)
\nsurf_{H_2}, 
\label{eq:rate thH2}
\end{eqnarray}
and
\begin{eqnarray}
\dot{\nsurf}_{H}&=&{\cal S}_H+\tau_{bsH}^{-1}
\nbulk_H(\fracsurf B^s/\fracbulk)-(\tau_{sbH}^{-1}B^b
+2\alpha_{Fs}S+\Gamma_H)\nsurf_H-2(\Gamma_{2H}+
t^{-1}_{Fs})\fracsurf^{-1}\nsurf_H^2\nonumber \\
&-&(t_{Fsb}^{-1}+t_{Fbs}^{-1})\fracbulk^{-1}\nbulk_H\nsurf_H+
(2\tilde{\tau}^{-1}_{Ds}B^s+t^{-1}_{Dsb}B^b)\nsurf_{H_2}
+(\fracsurf B^s/\fracbulk)t_{Dbs}^{-1}\nbulk_{H_2}.
\label{eq:rate thH}
\end{eqnarray}
In steady state $\dot{\nbulk}_H=\dot{\nbulk}_{H_2}=\dot{\nsurf}_{H_2}=
\dot{\nsurf}_{H}=0$.  Then the sum of eqs. (\ref{eq:rate NH}) 
and (\ref{eq:rate thH}) and twice eqs. (\ref{eq:rate NH2}) 
and (\ref{eq:rate thH2}) yields,
$$
S=S_H+{\cal S}_H= t^{-1}_{H-out} + 2t^{-1}_{H_2-out} 
$$
where the loss rate of H to the vacuum is
$$
t^{-1}_{H-out} = \Gamma_H\nsurf_H+
2B^s(\tau_{Ds}^{-1}-\tilde{\tau}_{Ds}^{-1})\nsurf_{H_2}
=\Gamma_H\nsurf_H+2q\beta_sSB^s\nsurf_{H_2},
$$
and the loss rate of H$_2$ to the vacuum is
$$
t^{-1}_{H_2-out} = \Gamma_{H_2}\nsurf_{H_2}+
\fracsurf^{-1}\Gamma_{2H}\nsurf_H^2
+\alpha_{Fs}S\nsurf_H(1-\delta_{Fs}).
$$
The efficiency of H$_2$ formation is the 
fraction of hydrogen released from the grain in molecular form:
\begin{equation}
{\cal E}={ 2 t^{-1}_{H_2-out} \over  t^{-1}_{H-out} + 2 t^{-1}_{H_2-out} }
= 1 - { t^{-1}_{H-out} \over S}.
\label{eq:effic}
\end{equation}

Let us examine the solution to the above equations in various limits.  
First, suppose that the timescales for H$_2$ formation and thermal 
dissociation on the surface, $t_{Fs}$ and $\tau_{Ds}$, 
are much shorter than any other timescale involved.  We then anticipate 
the H and H$_2$ to be in equilibrium on the surface.  Indeed, in the 
limit where all the rates except the formation
and dissociation rates tend to zero, equations (\ref{eq:rate thH2}) 
and (\ref{eq:rate thH}) yield, $B^s(\nsurf_{H_2}/\fracsurf)=
(\tau_{Ds}/t_{Fs})(\nsurf_H/\fracsurf)^2=
e^{2E_d^s/kT_g}(\nsurf_H/\fracsurf)^2$, 
which is just the appropriate mass action law (note
$E_d^s$ is the dissociation energy per atom).  Similarly, when H$_2$ 
formation and dissociation are the dominant processes in the 
bulk, eqs. (\ref{eq:rate NH}) and (\ref{eq:rate NH2}) yield, 
($\nbulk_{H_2}/\fracbulk)B^b=(\tau_{Db}/t_{Fb})(\nbulk_H/\fracbulk)^2=
e^{2E_d^b/kT_g}(\nbulk_H/\fracbulk)^2$.  Next, suppose that the surface (bulk)
to bulk (surface) diffusion times are very short compared with any 
other timescale. Upon taking the limit where all rates except the exchange 
rates tend to zero in eqs. (\ref{eq:rate NH}) and (\ref{eq:rate thH}), 
we recover the equilibrium distribution of H: $B^b(\nsurf_H/\fracsurf)=
B^s(t_{sbH}/t_{bsH})(\nbulk_H/\fracbulk)=
B^s\exp\{(E[H]^b-E[H]^s)/kT_g\}(\nbulk_H/\fracbulk)$.
Finally, in the limit where the H evaporation time is much shorter than 
quantum pair recombination, H$_2$ formation and dissociation on the 
surface, and transfer times from bulk to surface and 
vice versa, we obtain from eq. (\ref{eq:rate thH}), $\nsurf_H=
{\cal S}_H/\Gamma_H$, as expected.  Likewise, when the pair recombination 
rate dominates, eq. (\ref{eq:rate thH}) gives $\nsurf_H=
(\fracsurf{\cal S}_H/2\Gamma_{2H})^{1/2}$.

In the following, we shall neglect the densities in the statistical
factors, as well as the surface-subsurface reactions.  Specifically,
we set $B^s=B^b=1$,
$t_{Fsb}^{-1}=t_{Fbs}^{-1}=t_{Dsb}^{-1}=t_{Dbs}^{-1}=0$.  This
approximation enables us to solve eqs. (\ref{eq:rate NH}) through
(\ref{eq:rate thH}) analytically in various regimes.  The neglect of
the densities in the statistical factors is certainly justified when
the grain is unsaturated (i.e., $\nbulk_H + \nbulk_{H_2}<<\fracbulk$,
$\nsurf_H + \nsurf_{H_2}<<\fracsurf$), a condition we check in our
solutions.  The full numerical solution of eqs. (\ref{eq:rate
NH})-(\ref{eq:rate thH}) indicates that this is a good approximation
even for relatively large concentration numbers.  In ignoring cross
terms between the surface and bulk we are essentially assuming that the
mixed rates are not larger than the corresponding rates in both the surface and
the bulk. We have checked our results in a
variety of limits and find no qualitative difference for the parameter
ranges we have explored. 

\subsubsection{Equations without H$_2$ Formation/Destruction in the Bulk}

We now explore the conditions required for effective H$_2$ 
formation in different regimes.  First, consider a grain for which
neither formation nor destruction of H$_2$ is possible
in the bulk. This 
corresponds to the limit $\alpha_{Fb}=t_{Fb}^{-1}=t_{Db}^{-1} = \beta_b=0$ 
in eqs. (\ref{eq:rate NH}) - (\ref{eq:rate thH}).  
Then eq. (\ref{eq:rate NH}) can be written in the form,
\begin{equation}
\nbulk_H(a_1,a_2,a_3,a_4)=a_3\nsurf_H(a_1,a_2)+a_4,
\label{eq:sol NH}
\end{equation}
eq. (\ref{eq:rate NH2}),
\begin{equation}
\nbulk_{H_2}(a_1,a_2,a_5,a_6,a_7)=a_7\nsurf_{H_2}(a_1,a_2,a_5,a_6),
\label{eq:sol NH2}
\end{equation}
and eq. (\ref{eq:rate thH2}),
\begin{equation}
\nsurf_{H_2}(a_1,a_2,a_5,a_6)=a_5\nsurf_H^2(a_1,a_2)+a_6\nsurf_H(a_1,a_2).
\label{eq:sol thH2}
\end{equation}
The sum of eqs.(\ref{eq:rate NH})  and (\ref{eq:rate thH}) gives the 
following dimensionless equation for the surface occupation probability:
\begin{equation}
1-a_1\nsurf_H(a_1,a_2)-a_2\nsurf_H^2(a_1,a_2)=0.
\label{eq:sol thH}
\end{equation}
Here
\begin{eqnarray}
a_1&=& \Gamma_HS^{-1}+2\bar{\alpha}_{Fs};\ \ \ \ \bar{\alpha}_{Fs}=
\alpha_{Fs}\left(1-{\delta_{Fs}
\tilde{\tau}_{Ds}^{-1}\over \tau_{Ds}^{-1}+\Gamma_{H_2}}
\right),\nonumber \\ a_2&=&2\fracsurf^{-1}S^{-1}(\Gamma_{2H}+\bar{t}_{Fs}^{-1})
;\ \ \ \ \ \bar{t}_{Fs}^{-1}=t_{Fs}^{-1}\left(
1-{\tilde{\tau}_{Ds}^{-1}\over \tau_{Ds}^{-1}+\Gamma_{H_2}}\right),\nonumber \\
a_3&=&(\fracbulk/\fracsurf)(\tau_{bsH}/\tau_{sbH}), \nonumber \\
a_4&=&(\fracbulk/\fracsurf)\tau_{bsH}S_H,\nonumber\\
a_5&=&{\fracsurf^{-1}t_{Fs}^{-1}\over\tau_{Ds}^{-1}+\Gamma_{H_2}},\nonumber \\
a_6&=&{\alpha_{Fs}\delta_{Fs}S\over\tau_{Ds}^{-1}+\Gamma_{H_2}},\nonumber \\
a_7&=&(\fracbulk/\fracsurf)(\tau_{bsH_2}/\tau_{sbH_2}),
\label{eq:a coeff}
\end{eqnarray}
are dimensionless parameters that depend on the basic physical parameters as stated 
explicitly above.

As seen from eqs. (\ref{eq:sol thH}) and (\ref{eq:a coeff}), the
surface occupation number, $\nsurf_H$, depends on only two of the
seven dimensionless parameters of the problem. For example, only the
total injection rate ($S$) but not the individual bulk ($S_H$)
and surface (${\cal S}_H$) injection
rates are relevant for the value of $\nsurf_H$.  Since there is no
interconversion of H and H$_2$ inside the grain, in
steady state every atom intercepted by the grain must eventually
escape through the surface. We will exploit the dependence of
$\nsurf_H$ on $a_1$ and $a_2$ in the discussion in the next two 
sections. Note that
in the simplest case with no H$_2$ formation on the surface
($t_{Fs}^{-1} = 0$), no prompt reaction ($\alpha_{Fs} = 0$) and pair
evaporation only by quantum tunneling we have $-\ln a_1 \propto
-\ln \Gamma_H \propto E_e/kT_g$ and $-\ln a_2 \propto
-\ln \Gamma_{2H} \propto G_H$. The essential microphysical parameters
that govern the surface H density are $E_e$ and $G_H$. We will begin
by characterizing the solutions to the steady-state problem directly
in terms of possible combinations of $E_e$ and $G_H$.  We
will then argue that more general cases are directly and simply
described in terms of similar regions expressed as functions of $-\ln a_1$
and $-\ln a_2$, the analogues of $E_e$ and $G_H$.

Although the surface density depends solely on $a_1$ and $a_2$, all
the dimensionless parameters are needed to find a full solution. That
is important because we must check for saturation of surface and bulk
components as well as calculate ${\cal E}$.

\subsubsection{H$_2$ Formation by Pair Evaporation and/or Prompt Reaction}

To elucidate the role of pair evaporation and prompt reaction, 
we further assume a grain for which H$_2$ is unstable on the 
surface.  This corresponds to the limit $\tau_{Ds}\rightarrow0$, 
$\tau_{Ds}/\tilde{\tau}_{Ds}\rightarrow1$ in eq. (\ref{eq:a coeff}).  
Then, $a_2=2\fracsurf^{-1}\Gamma_{2H}/S$ and $a_5=a_6=0$.
Eqs. (\ref{eq:sol NH2}) and (\ref{eq:sol thH2}) then implies, 
$\nbulk_{H_2}=\nsurf_{H_2}=0$.  Now, when $a_4>\fracbulk$ the 
injection rate of H atoms into bulk sites, $S_H$, exceeds the 
rate at which hydrogen is released from the bulk, $\fracsurf
\tau^{-1}_{bsH}$, so the bulk will become saturated regardless of 
the values of the other parameters, as may be seen
from eq. (\ref{eq:sol NH}).  If $a_4<<\fracbulk$ bulk occupancy
is controlled essentially by H transfer from the surface to the 
bulk rather than injection of H in the bulk, that is, by the first 
term rather than the second
on the right hand side of eq. (\ref{eq:sol NH}).  
The grain will remain unsaturated if the evaporation, surface 
recombination or prompt reaction rates are sufficiently large.  
(The efficiency ${\cal E}=1-\Gamma_H\nsurf_H/S$ (see eq. 
[\ref{eq:effic}]) is also independent of $a_4$.) Provided we 
are not interested in the exact value of $\nbulk_H$ so long as 
it remains below $\fracbulk$ we can proceed by setting 
$a_4=0$ in eq. (\ref{eq:sol NH}).  The solution for $\nbulk_H$ is 
then exact in situations in which the ambient gas is cold, such 
that H atoms can only stick to the surface (i.e., $S_H=0$, 
${\cal S}_H=S$), and is a good 
approximation in general, provided that the rate at which H atoms are 
intercepted by the bulk is much smaller than the characteristic
rate of transfer from the bulk to the surface.

When $a_3<\fracbulk/\fracsurf$, or equivalently 
$\tau_{bsH}/\tau_{sbH}<1$, the steady-state distribution
of hydrogen in the grain always satisfies $\nbulk_H/\fracbulk
<\nsurf_H/\fracsurf$.  Consequently, when the surface reaches 
saturation the bulk remains unsaturated.  When  
$a_3>\fracbulk/\fracsurf$, bulk saturation always precedes surface 
saturation.  Let us define a critical surface concentration above which 
the grain (either bulk or surface) is saturated:
\begin{equation}
\nsurf_{sat}/\fracsurf=\left\{\begin{array}{lll}
	1; &\mbox{$\tau_{bsH}/\tau_{sbH}<1$} &\mbox{surface saturated}\\
	\tau_{sbH}/\tau_{bsH};&\mbox{$\tau_{bsH}/\tau_{sbH}>1$} &\mbox{bulk saturated}.
	\end{array}
\label{eqn:sat eqn}
\right.
\end{equation}
If thermal diffusion alone is responsible for surface-bulk fluxes,
then $\tau_{bsH}/\tau_{sbH} = t_{bsH}/t_{sbH} = e^{-(E_s-E_c)/kT_g}$
and the top line applies for $E_s > E_c $, the bottom for $E_s < E_c$.
The critical surface concentration has the simplest form
($\nsurf_{sat} = \fracsurf$) when $E_s > E_c$.  The condition for
either bulk or surface to be saturated is temperature dependent when
$E_s < E_c$ or whenever the rate coefficient includes collisionally
driven exchange. At several points in this paper we will make the
further assumption that the equations governing surface concentrations
continue to apply even after the bulk has become saturated. Formally,
the blocking factors that we have dropped would ensure mathematically
well-defined solutions. The critical assumption is really that the
basic grain parameters, measured or calculated for low occupancy
continue to remain accurate in the saturated limit.

On substituting $\nsurf_{sat}$ into eq. (\ref{eq:sol thH}), we obtain 
the equation of a surface in the space of physical parameters that 
separates the regimes of saturated and unsaturated grain.  
Specifically, the condition that the grain will be unsaturated reads,
\begin{equation}
{\Gamma_{2H} \over \nu_o}  >  \left( \fracsurf S 
\over 2 \nsurf_{sat} \nu_o \right)
\left( \nsurf_{sat}^{-1}-2\bar{\alpha}_{Fs}-
   {\Gamma_H \over S} \right)
\label{eq:NH sat1}
\end{equation}
with
\begin{equation}
{ \Gamma_H \over \nu_o} = \exp(-E_e/kT_g)+{\alpha_{svH}S\over \nu_o}
\end{equation}
where eq. (\ref{eq:sol thH}) together with the definitions of the
various physical parameters given above have been used. This
inequality may be viewed as a relation between $\Gamma_{2H}$ and
$E_e/kT_g$ when other parameters are fixed. In the absence of
collisional ejections and prompt reactions the grain will be
unsaturated either if $\Gamma_{2H} > \fracsurf S/2 \nsurf_{sat}^2$ or
if $E_e/kT_g < \ln ( \nu_o \nsurf_{sat}/S)$.

In a similar manner, we can derive an equation describing a family of 
isoefficiency surfaces, where the efficiency is defined in eq. 
(\ref{eq:effic}):
\begin{equation}
{ \Gamma_{2H} \over \nu_o } =
{ \fracsurf \Gamma_H \over 2 (1 - {\cal E})^2 S \nu_o }
\left( {\cal E} \Gamma_H - 2 (1 - {\cal E}) \bar{\alpha}_{Fs} S \right) .
\label{eq:eff cont1}
\end{equation}
Note that the isosurfaces are independent of the timescales 
$\tau_{sbH}$ and $\tau_{bsH}$ (and $\nsurf_{sat}$), 
as one might expect, since (i) evaporation 
and H$_2$ formation processes are restricted to the surface in the 
example presented here, and (ii) as already emphasized,
the surface concentration depends on the total injection rate alone.  In 
the absence of collisional ejections and prompt reactions eq. 
(\ref{eq:eff cont1}) assumes the simple form, $-\ln(\Gamma_{2H}/\nu_o)=
2E_e/kT_g+\ln[2S(1-{\cal E})^2/({\cal E}\fracsurf\nu_o)]$.  

Projections of the surfaces governed by eqs. (\ref{eq:NH sat1}) (when
the inequality sign is replaced by the equal sign) and (\ref{eq:eff
cont1}) onto the plane defined by $-\ln(\Gamma_{2H}/\nu_o)$ and
$E_e/kT_g$ are exhibited graphically in figure~\ref{fig:3}, for
$\fracsurf=10^{-2}$, and different choices of the remaining parameters
(note that the dependence of the curves on $\fracsurf$ as well as the
other physical parameters besides $-\ln(\Gamma_{2H}/\nu_o)$ and
$E_e/kT_g$ is logarithmic).  The regimes of saturated grain, effective
and ineffective H$_2$ formation are indicated.  The regions indicated
as effective and ineffective in the figures are separated by the
curves along which the efficiency ${\cal E}=1/2$.  In the regimes
indicated as effective the pair evaporation rate exceeds the
evaporation rate of H and in those indicated as ineffective the
opposite is true.  Windows (a) and (b) present a system not subject to
direct collisional processes ($\alpha_{svH}=\alpha_{Fs}=0$).  The
region enclosed by the solid line in case (a) corresponds to a
saturated surface, whereas the regions enclosed by the curves labeled
by values of $\tau_{bsH}/\tau_{sbH}>1$ correspond to bulk saturation.
Note that for any given substance,
$\nsurf_{sat}$ is variable, as discussed above.  (In a
later section we speculate upon the steady state of grains saturated
with hydrogen, and discuss several plausible scenarios for H$_2$
formation in such systems.)

It should be noted that $\Gamma_{2H}$ is constant only in the limit of
zero temperature quantum tunneling. The pair evaporation rate
generally depends on the grain temperature and $\Gamma_{2H} \to \nu_o$
as $T_g\rightarrow\infty$.  For $E_c>0$, the limiting form for pair
recombination (eq. [\ref{eq:pair-rec-limit1}]) implies
$-\ln(\Gamma_{2H}/\nu_o)\simeq (E_r/E_e)(E_e/kT_g)$ at grain
temperatures above the tunneling temperature $T_t$, and
$-\ln(\Gamma_{2H}/\nu_o)\simeq G_H$ below; for $E_c<0$ the limiting
form (eq. [\ref{eq:pair-rec-limit2}]) implies
$-\ln(\Gamma_{2H}/\nu_o)\simeq [(E_a-E_c)/E_e](E_e/kT_g)$ above, and
$-\ln(\Gamma_{2H}/\nu_o)\simeq G_H-E_c/kT_g$ below.  In figure~\ref{fig:4}, we
show the dependence of pair evaporation rate on grain temperature
($-\ln \Gamma_{2H}/\nu_o$ is plotted against $E_e/kT_g$) for two
materials.  The dashed line describes a typical transition metal with
$E_e=2.55$ eV, $E_c=-0.3$ eV, $E_a=0.2$ eV, and $G_H>7$.  The
dotted-dashed line describes carbon (with $E_c > 0$) using the same
parameters adopted in \S \ref{sec:app}, namely $E_e=0.3$ eV, $E_r=0.4$
eV, and $G_H=30$.  The solid lines for efficiency and
surface saturation were computed using the same parameters as in figure~\ref{fig:3}(a)
with $\tau_{bsH}/\tau_{sbH}<1$. Note that the conditions for bulk saturation
are not displayed in this figure.  In the case of carbon we find that 
bulk saturation occurs (for $E_s=-0.2$ eV, $S\simeq 10^{-13}$ s$^{-1}$, 
and the above choice of the remaining parameters, see
\S \ref{sec:app} for a detailed discussion) when $E_e/kT_g >7$.   
For both of these putative materials the surfaces remain unsaturated.
The threshold temperature below which effective H$_2$
formation takes place is $\sim 10^3$ K for the transition metal and
$\sim 90$ K for carbon.

The effect of collisional ejection of H from the surface and prompt
reaction is illustrated in figure~\ref{fig:3} windows (c) and (d).  The range of
values of $\alpha_{svH}$ and $\bar{\alpha}_{Fs}$ shown in these
windows encompasses most of the physically allowed range, namely from
zero to $\ntotal$ (in which case every incident atom 
induces a reaction).  In both cases the region labeled
saturated shrinks with increasing values of the efficiencies
$\alpha_{svH}$ and $\bar{\alpha}_{Fs}$, and eventually disappears when
$\alpha_{svH}+2\bar{\alpha}_{Fs}=\nsurf_{sat}^{-1}$ (see 
eq. [\ref{eq:NH sat1}]).  The reason is that both
processes involve release of hydrogen from the grain and,
consequently, when $\alpha_{svH}+ 2\bar{\alpha}_{Fs}=\nsurf_{sat}^{-1}$ the
grain can be sustained unsaturated irrespective of the thermal and
quantum release rates.  For example, the dotted and dash-dotted curves
in windows (c) and (d) correspond to a case where the grain is
unsaturated in the entire $(- \ln \Gamma_{2H}/\nu_0, E_e/kT_g )$ plane.
The panels show that
the H$_2$ formation efficiency decreases with
increasing $\alpha_{svH}$ and increases with increasing $\alpha_{Fs}$.
Because case (c) involves release of H atoms by direct
collisions, the efficiency curves become independent of $E_e$ for
values larger than that at which the collisional ejection rate exceeds
the thermal evaporation rate, that is, $\Gamma_H<\alpha_{svH}S$.
Likewise, case (d) involves collisional formation followed by ejection
of H$_2$ molecules and, therefore, the efficiency curves become
independent of $\Gamma_{2H}/\nu_0$ for values such that
$\Gamma_{2H}<\alpha_{Fs}S$.

\subsubsection{H$_2$ from Surface Formation plus 
Pair Evaporation and Prompt Reaction} 

We now generalize the above analysis to grains whose surfaces can
form and dissociate H$_2$ while continuing to assume no H$_2$
formation/dissociation in the bulk.  Consequently, the rate constants
$\tau_{Ds}^{-1}$, $t_{Fs}^{-1}$ and $\bar{t}_{Fs}^{-1}$ assume non-trivial values
and the definition of $\bar{\alpha}_{Fs}$ is modified. The
dimensionless coefficients $a_5$ and $a_6$ acquire
non-zero values but the previous analysis remains basically unchanged.
In particular, the rate coefficients $a_3$ and $a_4$ and the
expression for $\nsurf_{sat}$ remain unaltered.  
H saturation in the bulk still occurs when $a_4>\fracbulk$
irrespective of the values of the other parameters. 
The presence of H$_2$ formation on the surface does
not change the characteristic release rate of H from the bulk.  (This
would not be true in the presence of H$_2$ formation in the bulk,
because then hydrogen can be released from the bulk also in molecular
form.)  We proceed, as before, by setting $a_4=0$.  For $q=0$,
(no ejections of H atoms following collisional dissociation on
the surface) eq. (\ref{eq:NH sat1}), which gives the condition for H
not being saturated, and eq. (\ref{eq:eff cont1}), which gives
the efficiency, are modified as follows:
\begin{eqnarray}
\Gamma_{2H} & \to & \Gamma_{2H} + \bar{t}_{Fs}^{-1} \\
            & = & \Gamma_{2H} + { t_{Fs}^{-1} \Gamma_{H_2} \over
                                 \tau_{Ds}^{-1} + \Gamma_{H_2}}
\end{eqnarray} 
Since the rates of surface formation
and pair evaporation are each quadratic in $\nsurf_H$, the previous
analysis is only altered by the change in the effective H$_2$
formation rate coefficient.  The indicated change above clearly shows
that the efficiency will increase when H$_2$ can form on the surface
but the increase is qualitatively important only when pair evaporation
and prompt reactions are ineffective by themselves ($S
(\Gamma_{2H}/\fracsurf \Gamma_H^2 + 2 \alpha_{Fs}/\Gamma_H) << 1$)
while the additional process is sufficiently rapid ($S
\bar{t}_{Fs}^{-1}/\fracsurf \Gamma_H^2 \simgreat 1$). The form of
$\bar{t}_{Fs}^{-1}$ depends on the ratio of surface dissociation rate
to evaporation rate for H$_2$. If evaporation is
more rapid then $\bar{t}_{Fs}^{-1} \to t_{Fs}^{-1}$ and nearly every
surface-formed H$_2$ escapes as soon as it is created. Efficient
molecule formation requires $S t_{Fs}^{-1}/\fracsurf \Gamma_H^2
\simgreat 1$.  On the other hand, if dissociation
is more rapid then H and H$_2$ obey a Saha-like equilibrium on the
surface and the molecule escape rate is the evaporation rate of the
H$_2$ fraction.  Efficient formation requires $S
\Gamma_{H_2} e^{2E_d^s/kT_g}/\fracsurf \Gamma_H^2 \simgreat 1$.

The formation of H$_2$ on the surface provides a source of molecular hydrogen 
on and in the grain and one must worry about the possibility of H$_2$ 
saturation in addition to H saturation.  The region in the
space spanned by the parameters $a_1$ through $a_7$ in which the grain is 
unsaturated by any component is determined, in terms of the solution of 
eqs. (\ref{eq:sol NH})-(\ref{eq:sol thH}), by the set of inequalities,
\begin{eqnarray}
\nsurf_H(a_1,a_2)+\nsurf_{H_2}(a_1,a_2,a_5,a_6) & < & \fracsurf \nonumber \\
\nbulk_H(a_1,a_2,a_3,a_4)+\nbulk_{H_2}(a_1,a_2,a_5,a_6,a_7) & < & \fracbulk. 
\end{eqnarray}
This also defines the regime of unsaturated grain in the space of physical 
parameters through the dependence of $a_1$-$a_7$ on them.

Rather surprisingly, it is possible to give an approximate compact
analytic treatment of the above conditions. We have the limiting
solutions for $\nsurf_H$
\begin{equation}
\nsurf_H = 
	\left\{ 
	\begin{array}{ll}
		1/a_1 & \mbox{for $a_1^2 >> 4a_2$} \\
		1/a_2^{1/2} & \mbox{for $a_1^2 << 4a_2$}
	\end{array}
	\right.
\end{equation}
and the condition that the grain (both surface and bulk) be
unsaturated is
\begin{equation}
\nsurf_H^2 <<
   \min\left(
{\fracsurf \over a_5},
\left( \fracsurf \over 1 + a_6 \right)^2,
{\fracbulk - a_4 \over a_5 a_7},
\left( \fracbulk - a_4 \over a_3 + a_6 a_7 \right)^2
\right)
\label{eq:sat H H2}
\end{equation}
Terms involving $\fracsurf$ ($\fracbulk$) are the surface (bulk)
constraints.  (As discussed previously, 
if $\fracbulk < a_4$ the bulk must become saturated.)
These inequalities guarantee that the regions of saturated and
unsaturated grains will be identical in shape to those shown in window
(a) of figure~\ref{fig:3} in which the axes are relabeled, viz.
\begin{eqnarray}
{E_e \over kT_g} \to -\ln {S a_1 \over \nu_o} \nonumber \\
{\Gamma_{2H}} \to {\fracsurf S a_2 \over 2}
\end{eqnarray}
and the numerical values for the horizontal and vertical
lines are determined from the above list.

It is also straightforward to show that the limiting forms
for efficiency for $q=0$ are
\begin{equation}
{\cal E} = 
	\left\{
\begin{array}{ll}
\left( 2 \bar{\alpha}_{Fs} S \over \Gamma_H + 2 \bar{\alpha}_{Fs} S \right)
& \mbox{for $a_1^2 >> 4a_2$} \\
1 - \left( \fracsurf \Gamma_H^2 \over 2 S ( \Gamma_{2H} + \bar{t}_{Fs}^{-1} ) 
    \right)^{1/2}
& \mbox{for $a_1^2 << 4a_2$}
\end{array}
\right. .
\end{equation}
When $q \ne 0$, the efficiency is lowered by the amount $2 q \beta_s
\nsurf_{H_2}$.

\subsubsection{H$_2$ Formation in the Bulk}

Next, we wish to examine the conditions under which effective H$_2$
formation may take place in the bulk, and explore the different
regimes in parameter space in which grain saturation by H and/or H$_2$
occurs.  Since the inclusion of quantum pair evaporation complicates
the analysis considerably, and can only lead to a higher H$_2$
formation efficiency anyway, we ignore it in the present example.  We
also ignore H$_2$ formation and dissociation on the surface (but keep
the prompt reaction process) in order to simplify the analysis
further.  The complementary case in which H$_2$ formation is controlled
entirely by the surface has been studied in the previous example.
			    
Consider a grain for which H$_2$ formation in the bulk is allowed but 
H$_2$ formation and destruction on the surface as well as quantum 
pair evaporation are negligible.  We set $t^{-1}_{Fs}
=\delta_{Fs}=\tau_{Ds}^{-1}=\Gamma_{2H}=0$ 
in eqs. (\ref{eq:rate NH})-(\ref{eq:rate thH}).  One then finds,
\begin{equation}
1- b_1\nbulk_H - b_2 \nbulk_H^2=0,
\label{eq:sol NHb}
\end{equation}
with
\begin{eqnarray}
b_1 & = & \left( {2 \alpha_{Fb} \over 1 + \eta} + 
{\left( \fracsurf \tau_{sbH} \over \fracbulk \tau_{bsH} \right)}
          {\Gamma_H/S + 2 \alpha_{Fs} \over
           1 + \tau_{sbH} (\Gamma_H + 2 \alpha_{Fs}S) }
\right) r \nonumber \\
b_2 & = & {2\over \fracbulk t_{Fb} S (1 + \eta)} r
\label{eq:coeff b1b2}
\end{eqnarray}
and we have introduced two auxiliary variables to simplify the notation:
\begin{eqnarray}
\eta & = &{\fracbulk(1+\tau_{sbH_2}\Gamma_{H_2})\tau_{bsH_2}\over \fracsurf\tau_{sbH_2}
\Gamma_{H_2}\tau_{Db}} \nonumber \\
{1 \over r} & = & { {S_H \over S} + { {\cal S}_H \over S (1 + \tau_{sbH}(\Gamma_H + 2 \alpha_{Fs} S)) } } .
\end{eqnarray}
It can be easily verified that in the absence of H$_2$ formation 
inside the grain, which corresponds to the limit $t_{Fb}^{-1}=
\alpha_{Fb}=0$, the solution of eq. (\ref{eq:sol NHb}), given
by $\nbulk_H=1/b_1$, coincides with the solution of eq. (\ref{eq:sol
NH}) and (\ref{eq:sol thH}) for $\Gamma_{2H}=0, \tau_{Ds}=0$ as
required.  Note that the quadratic coefficient $b_2 \ne 0$ only if
H$_2$ can form by thermal processes within the bulk and its effect is
to lower $\nbulk_H$. The complexity of
the linear coefficient $b_1$ arises from all the linear processes
that alter the distribution of H (and H$_2$) in and on the grain
including evaporation from the surface ($\Gamma_H$, $\Gamma_{H_2}$), 
transfer between bulk and surface ($\tau_{sbH}$, $\tau_{bsH}$,
$\tau_{sbH_2}$, $\tau_{bsH_2}$), 
and collisional H$_2$ formation by particles incident on the
bulk and on the surface ($\alpha_{Fb}$, $\alpha_{Fs}$).

The quantity $1/r$ is the equilibrium fraction of the incident H atoms
available to the bulk. $r$ lies in the range between $1$, for
$S_H\simeq S$ (as in the case of homogeneous injection for which
${\cal S}_H = \fracsurf S$ and $S_H=\fracbulk S\simeq S$, provided
that the grain is large enough), and
$1+\tau_{sbH}(\Gamma_{H}+2\alpha_{Fs}S)$ for $S_H=0$ (which
corresponds to a situation where the ambient gas is cold such that
atoms impinging on the grain stick to the surface and diffuse into
the bulk).  In the
latter case, $\nbulk_H$ approaches zero when the surface to bulk
exchange time $\tau_{sbH}$ becomes much longer than the evaporation
and prompt reaction timescales since the only source of
bulk atomic hydrogen is the flux from the surface.

The coefficient $\eta$ is the ratio of dissociation rate and net
release rate of molecular hydrogen in the bulk.  If $\tau_{Db}$ is
much longer than the H$_2$ release time, viz., $\eta<<1$, almost every newly
formed molecule will escape from the bulk before it
dissociates and, consequently, the loss rate of bulk atomic
hydrogen will depend on the molecule formation rate (thermally
activated and collisional) alone.  When $\eta>>1$, only a fraction
$\eta^{-1}$ of the newly formed molecules leave the bulk before they
dissociate, so the loss rate of H in the bulk is reduced by a factor
$\eta^{-1}$.

The surface occupation probability of H satisfies the equation,
\begin{equation}
\nsurf_H=b_3\nbulk_H+b_4,
\label{eq:sol thHb}
\end{equation}
where eqs. (\ref{eq:rate thH2}) and (\ref{eq:rate thH}) have been 
employed, and where 
\begin{eqnarray}
b_3&=&{\fracsurf(\tau_{sbH}/\tau_{bsH})\over \fracbulk[1+\tau_{sbH}
(\Gamma_{H}+2\alpha_{Fs}S)]},\nonumber \\
b_4&=&{\tau_{sbH}{\cal S}_H\over 1+\tau_{sbH}(\Gamma_H+2\alpha_{Fs}S)}.
\label{eq:coeff b3b4}
\end{eqnarray}
The parameter $b_3$ represents the net contribution
from the bulk to surface by exchange,
whereas $b_4$ represents the contribution from direct injection 
of H atoms on the surface.  When $\tau_{sbH}(\Gamma_H+2\alpha_{Fs}S)>>1$, 
the loss of hydrogen atoms from the surface is dominated by
prompt reaction, evaporation and  direct ejections rather than transfer 
to the bulk.  Then $(\Gamma_H+2\alpha_{Fs}S)\nsurf_H\simeq 
\fracsurf\tau_{bsH}^{-1}(\nbulk_H/\fracbulk)+{\cal S}_H$; that is,
the loss of surface H atoms by evaporation, direct ejection, and 
prompt reaction is balanced by injection into surface sites plus 
transfer of H atoms from the bulk. 
In the opposite limit, $(\nsurf_H/\fracsurf)\simeq (\tau_{sbH}/\tau_{bsH})(\nbulk_H/\fracbulk)+
\tau_{sbH}{\cal S}_H/\fracsurf$,
which for sufficiently small surface injection rate 
($\tau_{sbH}{\cal S}_H\simeq0$) approaches the equilibrium distribution 
(see discussion following eq. [\ref{eq:rate thH}]).

The bulk and surface concentrations of molecular hydrogen depend
on three additional parameters:
\begin{eqnarray}
b_5&=&{\tau_{Db}\over \fracbulk t_{Fb}} \left( \eta \over 1 + \eta \right) ,\nonumber \\
b_6&=&\alpha_{F_b}S\tau_{Db} \left( \eta \over 1 + \eta \right) ,\nonumber \\
b_7&=&{1 \over \eta \tau_{Db} \Gamma_{H_2}}.
\label{eq:coeff b5b7}
\end{eqnarray}
In terms of these parameters, the solution for $\nbulk_{H_2}$ takes the simple form 
\begin{equation}
\nbulk_{H_2}=b_5\nbulk_H^2+b_6\nbulk_H,
\label{eq:sol NH2b}
\end{equation}
and that for $\nsurf_{H_2}$,
\begin{equation}
\nsurf_{H_2}=b_7\nbulk_{H_2}.
\label{eq:sol thH2b}
\end{equation}
The last equation reflects the fact that the only source of H$_2$ 
molecules on the surface is the flux of H$_2$ from the bulk.  This 
is due to our neglect of H$_2$ formation on the surface.

In general, the regime of validity of the solution presented above is defined by:
\begin{eqnarray}
\nbulk_H(b_1,b_2)+\nbulk_{H_2}(b_1,b_2,b_5,b_6)<\fracbulk,\nonumber\\
\nsurf_H(b_1,b_2,b_3,b_4)+\nsurf_{H_2}(b_1,b_2,b_5,b_6,b_7) <\fracsurf.
\label{eq:sat condb}
\end{eqnarray}
We now note that the substitution $b_i \to a_i$, the interchanges
$\nbulk \leftrightarrow \nsurf$ and
$\fracbulk \leftrightarrow \fracsurf$
yield exactly the same set of equations
analyzed in the preceding section. Thus, we have the limiting
solution $\nbulk_H$
\begin{equation}
\nbulk_H = 
	\left\{ 
	\begin{array}{ll}
		1/b_1 & \mbox{for $b_1^2 >> 4b_2$} \\
		1/b_2^{1/2} & \mbox{for $b_1^2 << 4b_2$}
	\end{array}
	\right.
\end{equation}
and the condition that the grain (both surface and bulk) be
unsaturated is
\begin{equation}
\nbulk_H^2 <<
   \min\left(
{\fracbulk \over b_5},
\left( \fracbulk \over 1 + b_6 \right)^2,
{\fracsurf - b_4 \over b_5 b_7},
\left( \fracsurf - b_4 \over b_3 + b_6 b_7 \right)^2
\right)
\end{equation}
When $b_4>\fracsurf$, the injection rate of hydrogen on the surface
exceeds the rate at which H atoms are lost from the surface by
evaporation, prompt reaction, and transfer to the bulk, as can be seen
from eq.  (\ref{eq:coeff b3b4}), and, therefore, surface saturation
will occur regardless of the value of $\nbulk_H$.

In contrast to the problem of H$_2$ formation studied in the previous
section, there is no simple relationship between $b_1$ and $b_2$ and 
the fundamental microphysical parameters (e.g. $E_e/kT_g$ and $G_H$).
This is already apparent in the original expressions for $b_1$ and
$b_2$ which involve complex relationships between all the formation
and release processes for H and H$_2$. Nonetheless, the above
inequalities faithfully convey the conditions for surface and
bulk to remain unsaturated.

We define the efficiency of H$_2$ formation as in the previous example: 
${\cal E}=1-\Gamma_H\nsurf_H/S$.
By employing eqs. (\ref{eq:sol NHb}) and (\ref{eq:sol thHb}), we can 
express ${\cal E}$ as
\begin{equation}
{\cal E}=1-\Gamma_HS^{-1}[b_3\nbulk_H(b_1,b_2)+b_4]=1-\Gamma_HS^{-1}
\left\{{b_3\over 2 b_2}\left[-b_1+
\sqrt{b_1^2+4b_2}\right]+b_4\right\}.
\label{eq:effic b}
\end{equation}
The efficiency is independent of the 
parameters $b_5$ through $b_7$, as it measures only the ratio of escape 
rate (thermal evaporation plus direct ejections)
to interception rate by the grain of H nuclei.  It does not tell us 
anything about the values of bulk and surface
concentrations of molecular hydrogen.  The latter are determined by 
the rate coefficients $b_5$, $b_6$ and $b_7$, which involve ratios of 
H$_2$ formation rates to dissociation, release, and bulk to surface 
exchange rates, as seen from eq. (\ref{eq:coeff b5b7}).

To elucidate the solution further we consider the limit 
$\alpha_{Fs}=0$ and ${\cal S}_H = 0$ ($r = 1$).
The latter approximation implies $b_4 = 0$. We define
\begin{eqnarray}
x_F&=&1/t_{Fb}S,\nonumber \\
x_R&=&{\fracsurf(\tau_{sbH}/\tau_{bsH})(\Gamma_{H}/S)\over 
\fracbulk(1+\tau_{sbH}\Gamma_{H})} . 
\label{eq:var b}
\end{eqnarray}
$x_F$ is the thermally activated H$_2$ formation rate 
in the bulk in units of the injection rate per site, $S$.
$x_R$ is the net release rate per bulk H atom of atomic 
hydrogen from the bulk measured in units of $S$.  For 
$\tau_{sbH}\Gamma_H>>1$, $x_R\simeq(\fracsurf/\fracbulk) 
(\tau_{bsH}S)^{-1}$, in which case H release from the bulk is 
limited by the rate at which H is transferred to
the surface.  For $\tau_{sbH}\Gamma_H<<1$, $x_R\simeq 
(\fracsurf/\fracbulk)(\tau_{sbH}/\tau_{bsH})(\Gamma_H/S)$
and the net release of H is limited by evaporation from the grain 
($\fracsurf \tau_{sbH}/\fracbulk \tau_{bsH}$ is the equilibrium ratio 
of surface and bulk H concentrations).  

Eq. (\ref{eq:sol NHb}) may be recast as a relation between the
formation rate and the release rate for fixed $\nbulk_H$
\begin{equation}
x_R  =  \left(\nbulk_H^{-1}-{2\alpha_{Fb}\over 1+\eta}\right)-{
2\nbulk_H\over \fracbulk(1+\eta)}x_F .
\label{eq:cont NH b}
\end{equation}
Likewise eq. (\ref{eq:effic b}) may be used to find contours
of fixed efficiency in the ($x_R,x_F$) plane
\begin{equation}
x_F  =  {\fracbulk[{\cal E}(1+\eta)x_R^2-2(
1-{\cal E})\alpha_{Fb}x_R]\over2(1-{\cal E})^2}.
\label{eq:cont eff b}
\end{equation}
Finally, once $b_4$, $b_5$, $b_6$, and $b_7$ are fixed, the critical
value of $\nbulk_H$ for which the grain becomes saturated in or on can
be computed from eq. (\ref{eq:sat condb}).
 
Contours of $\nbulk_H$ in the $(x_R,x_F)$ plane are exhibited
graphically in figure~\ref{fig:5} (values of $\nbulk_H$ label the curves), for
$\alpha_{Fb}=0$, $\eta=0$ (solid lines) and $\eta=10^2$ (dotted
lines). When the release rate is large, then $\nbulk_H$ is small
(near the top). When the rate is small and the formation rate is small,
then $\nbulk_H$ is large (bottom right); while if the formation rate is large,
then it is small (bottom left). Another way to view the results is
as follows. The vertical portion of the curve
(independent of $x_R$) corresponds to the regime where $\nbulk_H$ is
limited by H$_2$ formation; the horizontal portion (independent of
$x_F$) is the regime where $\nbulk_H$ is limited by release from the
bulk.  As $\eta$ increases the curves move leftward.  The reason, as
already explained above, is that for $\eta>1$ the rate at which
hydrogen leaves the bulk in molecular form diminishes by $\eta^{-1}$
and, therefore, shorter formation time (i.e., larger $x_F$) is
required to maintain $\nbulk_H$ at the same value, as also clearly
seen from eq.  (\ref{eq:cont NH b}).  Also shown schematically in figure
\ref{fig:5} is the curve separating the regimes of saturated and unsaturated
grain (dashed line).  The location of this curve depends on the values
of the remaining parameters ($b_4$ - $b_7$) through eq. (\ref{eq:sat
condb}).  The corresponding critical value of $\nbulk_H$ can range
between zero for $b_4\ge \fracsurf$ and $\sim \fracbulk$. 

The quantitative range of $\nbulk_H$ is easily derived:
For a given injection rate, $\nbulk_H$ cannot exceed
$\nbulk_o=(1+\eta)/2\alpha_{Fb}$, when capture of H atoms in the bulk
is balanced by release of H$_2$ molecules produced by direct
collisions.  For this value eq. (\ref{eq:cont NH b}) yields $x_R=0$,
$x_F=0$.  Sustaining $\nbulk_H$ at values smaller than $\nbulk_o$
requires, in addition to collisional formation of H$_2$ molecules,
either thermally activated H$_2$ formation ($x_F>0$), or H release
($x_R>0$).  For $\nbulk_H<<\nbulk_o$ the latter processes dominate
over collisional H$_2$ formation which ultimately becomes negligible.

Labeled efficiency contours are presented in figure~\ref{fig:6} for $\eta=0$, 
$\alpha_{Fb}=0$ (solid lines) and $\alpha_{Fb}=
10^4$ (dashed lines).  As seen, for large $x_R$ (large release rate) 
the efficiency curves are power laws with slope 2, reflecting the 
dominance of thermally activated over collisional H$_2$ formation.  As 
$x_R$ approaches $2(1-{\cal E})\alpha_{Fb}/{\cal E}$, the desired 
efficiency can be achieved by collisional H$_2$ formation alone 
and, hence, $x_F\rightarrow 0$.

\subsubsection{Saturated Grain}
What happens when the grain becomes saturated by either atoms or 
molecules?  Presumably, the saturation of the grain affects the 
embedding energies and diffusion times considerably.  
Unfortunately, the dependence of those parameters on filling 
factors is poorly understood.  Nevertheless, we can
still draw some general conclusions.  Consider a 
clean grain having a temperature $T_g$, initially devoid of H 
atoms and H$_2$ molecules, embedded in a gas of hydrogen atoms.  
Suppose first that molecular hydrogen formation in and on the 
grain is energetically unfavorable and/or inefficient (i.e., $t_{Fi}
\rightarrow \infty$) and that the mean evaporation time of 
an H atom, the prompt-reaction time and the mean surface recombination
time are all much longer than $t_{in}$.  Then after a time $\sim
\ntotal t_{in}$ the grain will become saturated by H atoms.  The
evolution of the system from this point on will depend on the modified
parameters.  If the embedding energy of H atoms is not altered
significantly, then a newly incident atom will either eject bound H
atoms from the grain as a result of collisions provided that its
energy is sufficiently large, push bulk atoms to the surface, or will
stop inside the grain and then diffuse to the surface and stick.  If
the yield for the first process is much smaller than unity, then
layers of atomic hydrogen should form on top of the original grain
surface.  Our expectation is that this will lead to a dramatic
increase of the surface recombination rate and, consequently,
effective H$_2$ formation, since the atom-surface binding energy
declines sharply with distance from the original grain surface.

Next, consider the possibility that molecular hydrogen formation in 
or on the grain is effective, and that the mean evaporation time of 
a molecule is much longer than $2t_{in}$.  The grain will become 
saturated by H$_2$ molecules after a time $\sim2\ntotal t_{in}$.  A newly 
injected atom will then either break up a bound H$_2$, ejecting one 
of the H atoms from the grain in the process, or be pushed from 
the bulk to the surface and then evaporate or stick, 
depending on the modified surface binding energy.  If 
it sticks to the surface for a time longer than about $t_{in}$, it 
will eventually recombine with a second injected atom to form a 
molecule (under astrophysical conditions, the surface migration time is 
anticipated to be much shorter than $t_{in}$). This will lead to the 
growth of layers of molecular hydrogen until the evaporation rate of 
atoms or molecules exceeds the value required for steady-state. 

Alternatively, it could be that when the filling factor becomes 
sufficiently large, a few per cent say, the embedding energies of 
H and H$_2$ are sufficiently reduced to allow the evaporation rate 
of either atoms or molecules to balance the flux of incident particles. 
The efficiency of molecular formation in this scenario would depend on
the value of the modified parameters at steady-state.
Our analysis should be applicable provided that the hydrogen has not 
undergone a phase transition to some ordered phase at these densities.
	
\section{Application: H$_2$ Formation in Carbon}
\label{sec:app}

As an example we consider H$_2$ formation in and on carbon grains.
For simplicity, we take the distribution of H atoms in the gas phase 
to be isotropic and monoenergetic with respect to the grain, and denote 
by $n({\rm H})$ and $\epsilon$ the number density and energy (measured in 
units of eV) of the H atoms.  The velocity of an H atom relative to 
the grain is then $v=1.4\times10^6 (\epsilon/\eV)^{1/2}$ cm s$^{-1}$, and 
the flux impinging on the grain is $n({\rm H})v\simeq4\times10^5 
(\epsilon/\eV)^{1/2} n({\rm H})$ cm$^{-2}$ s$^{-1}$.  In terms of the transmission
coefficient $T(\epsilon)$, defined in the text preceding 
eq. (\ref{eq:mean-rate}), and the characteristic grain size 
$a=a_{-5}10^{-5}$ cm, we can approximate the injection rate as
\begin{equation}
t_{in}^{-1}\simeq 3\times10^{-4} n({\rm H})(\epsilon/\eV)^{1/2}
a^2_{-5}T(\epsilon)\ \ {\rm s}^{-1},
\label{eq:t_in}
\end{equation}
where the grain's geometrical cross section has been taken to be of 
order $4\pi a^2$ (ignoring possible focusing effects).  Let $d_{-8}$ 
be the average interatomic separation of C atoms in \AA.  Then 
the total number of sites is roughly $\ntotal \simeq 4\times10^9 
(a_{-5}/d_{-8})^3$ and the relative number of surface sites is of order 
$\fracsurf\simeq2\times10^{-3} (a_{-5}/d_{-8})^{-1}$.  Thus,
\begin{equation}
S=(\ntotal t_{in})^{-1}\simeq 7.5\times10^{-14} n({\rm H})(\epsilon/\eV)^{1/2}a^{-1}_{-5}d_{-8}^{3}
T(\epsilon)\ \ {\rm s}^{-1}.
\label{eq:S-ex}
\end{equation}

Hydrogen adsorption on carbon takes place in different types of sites 
with different activation energies. The experimental values of the 
activation energies for adsorption lie between about 0.3 
and 2.2 eV \cite{Ca,Bar}.  The heat of adsorption (or equivalently 
the embedding energy $E_e$, see \S \ref{subsec:E-Lev} and figure~\ref{fig:1}) has 
not been determined experimentally yet.  However, 
several attempts have been 
made to calculate it for the hydrogen graphite system using quantum 
chemical codes \cite{Ar,Kl}.  The results of such calculations are 
highly uncertain, but seem to suggest the following: i) the most favorable 
site for chemisorption is the one which corresponds to the H atom being 
directly above a C atom.  The embedding energy 
lies in the range between $\sim 0.3$ and 1.14 eV, implying $E_c$ 
between 1.94 and 1.1 eV.  ii) the activation energy for H pair 
recombination ($E_r$ in figure~\ref{fig:1}) is about 0.4 eV.  As mentioned 
in \S \ref{subsec:E-Lev}, the detection of highly excited H$_2$ 
molecules formed on carbon surfaces at low temperatures led Schermann 
et al. \cite{Sc} to conclude that the embedding energy should not exceed 
$\sim$ 0.3 eV, consistent with the results of ref. \cite{Kl}. 
For $E_e$ in the range $0.3-1.1$ eV and $E_r=0.4$ eV, we obtain an 
activation energy of chemisorption in the range between 2.34 (note that 
this implies that there should be an activation barrier of $\sim 0.1$
eV for adsorption of atomic hydrogen) and 1.5 eV, respectively, consistent 
with the experimental values mentioned above.	
              
Taking $E_r=0.4$ eV, eq. (\ref{eq:pent. fac.}) yields for the penetration 
factor at zero temperature $G_H\simeq 30 (z_o/\AA)$, and eq. 
(\ref{eq:pair-rec}) gives $T_t\simeq 150 (z_o/\AA)$ K for 
the tunneling temperature, where $z_o$ is again the width of the barrier.
In the absence of collisional processes eq. (\ref{eq:NH sat1}) 
with $\nsurf_{sat}=\fracsurf$ implies that the surface will be 
maintained unsaturated at $T_g=0$ provided that
\begin{equation}
\ln[n({\rm H})(\epsilon/\eV)^{1/2}d_{-8}^{2}T(\epsilon)]<54-30(z_o/\AA).
\label{eq:unsat-ex}
\end{equation}
As indicated above, we are ignoring the possibility that bulk saturation
alters fundamental grain parameters and assume
this condition to suffice even if the bulk is saturated.
(We show that bulk saturation is likely below.)
Adopting $z_o=1 \AA$, $d=3 \AA$, $T(\epsilon)\simeq1$ for illustration, we 
find that quantum pair evaporation will keep the surface unsaturated 
as long as $n({\rm H})(\epsilon/\eV)^{1/2}<2.5\times10^{9}$ cm$^{-3}$.
Increasing the grain temperature will result of course in a smaller 
surface concentration.  Note that the condition (\ref{eq:unsat-ex}) 
is independent of the grain size since
both the incoming flux of H atoms and the outgoing flux of H pairs are 
proportional to the surface area.  

If molecules form {\it only} by pair evaporation from the surface, the
efficiency is given by eq. (\ref{eq:eff cont1}).  Setting the
collisional rates to zero and taking ${\cal E}=1/2$ in eq.
(\ref{eq:eff cont1}), we find that for the same choice of $z_o$, $d$
and $T(\epsilon)$, the efficiency will exceed 1/2 if
\begin{equation}
T_g<T_{crt}={10^4(E_e/1 {\rm eV})\over 40-\ln(n({\rm H})
(\epsilon/\eV)^{1/2})/2}\ \ {\rm K}.
\label{eq:Tc-ex}
\end{equation}
For example, for the value of $E_e$ inferred by Schermann et al., viz., 
$E_e=0.3$ eV, the last equation gives $T_{crt}$ between about 75 and 
100 K for $n({\rm H})(\epsilon/\eV)^{1/2}$ between 1 and $10^{7}$ 
cm$^{-3}$, respectively.  It 
is important to note that the efficiency ${\cal E}$ defined in
eq. (\ref{eq:effic}) is essentially the probability that an H atom 
intercepted by the grain will be released as part of a molecule.  The 
overall H$_2$ formation efficiency is the product of
${\cal E}$ and the transmission (sticking or penetrating) probability of H.
											                 
If collisional displacement of H from bulk to 
surface does not occur and if
molecule formation in the bulk does not occur
then we find that
the bulk will be saturated by H in steady state. Our reasoning
follows.  The solubility and diffusivity of hydrogen in the bulk are
uncertain.  The reported values of the diffusion energy range between
0.5 and 4.3 eV \cite{Wi,Ca}.  This is most likely due to the broad 
spectrum of binding sites.  The deepest sites are the trapping
sites thought to be associated with unsaturated carbon atoms on the 
edges of microcrystals constituting the amorphous material \cite{Ca,Kl}.  
The trap energy is of order 4.3 eV, comparable with the bond energy 
between H and C in typical hydrocarbons.  The mobility of hydrogen at 
low concentrations is presumably controlled by the deep sites.  In fact, 
implantation experiments appear to suggest that atomic hydrogen may 
be highly mobile in the absence of lattice damages 
\cite{HFV}.  Astrophysical grains are likely to be damaged.  However,
once the deep sites are filled the mobility of additional atomic
hydrogen in the bulk could be large.  The best value for the solution
energy, according to ref. \cite{Ca}, is that given by Atsumi et al.
\cite{ATM}: $E_s=-0.2$ eV (cf. figure~\ref{fig:1} case iv), and for the diffusion
energy is $E_D=2.8$ eV.  The energy difference between the surface and
bulk ground states is given by $E_e-E[H]^b=E_c-E_s$ (cf. \S 
\ref{subsec:E-Lev}).  With the values of $E_c$ discussed above, 
$E_s=-0.2$ eV implies that
this energy gap lies in the range between 1.3 and 2.1 eV.
The condition on the maximum equilibrium
abundance to prevent saturating the bulk is
very strict on account of the large energy gap:
$\nsurf_{sat} = \fracsurf e^{-(E_c-E_s)/kT_g}$ (cf. eq. [\ref{eqn:sat eqn}]).
If molecule formation on the surface is efficient we require
\begin{equation}
\ln[n({\rm H})(\epsilon/\eV)^{1/2}d_{-8}^{2}T(\epsilon)]<54
-30(z_o/\AA) \min \left( {T_t \over T_g}, 1 \right)
-{3\times10^4 \over T_g} \left( E_c - E_s \over 1.3 \eV \right),
\end{equation}
or if H evaporation dominates
\begin{equation}
\ln[n({\rm H})(\epsilon/\eV)^{1/2}d_{-8}^{2}T(\epsilon)]<54
-{2.8\times10^4 \over T_g} \left( E_e + E_c - E_s \over 2.44 \eV \right) .
\end{equation}
Neither condition can be satisfied for typical astrophysical conditions
so H saturation is favored.

It is important to recognize that the surface to bulk exchange rate
has a large typical activation energy $\Delta E=E_D+E_s-E_c$.  For
$E_D=2.8$ eV, we obtain $\Delta E$ between 1.5 and 0.7 eV.  Given
these estimates we anticipate the surface to bulk exchange rate to be
much smaller than H or quantum pair evaporation rate. Bulk saturation
is achieved, however, if high energy particles penetrate the lattice.

This relatively simple picture for graphite could be complicated by
the following additional chemical pathways.  The bulk may avoid
saturation in steady state if sufficiently rapid H$_2$ formation can
occur in the bulk followed by rapid bulk to surface exchange of H$_2$,
or in the presence of rapid collisional transfer of bulk H to the
surface.

Whether H$_2$ formation can take place in the bulk is unclear.  It
can be argued, based on the foregoing discussion, that if H$_2$ is
stable in the bulk then the activation barrier for bulk to surface
exchange of H$_2$ molecules is large (it must exceed that for H atoms;
c.f figure~\ref{fig:2}) so that H$_2$ saturation is inevitable for the same reasons
that H saturation is. In both cases, saturation of the deep binding
sites could alter the effective interaction potential.  By simply
occupying the deepest sites, hydrogen within the grain may force newly
added atoms or molecules to move amongst more weakly bound sites.
Alternatively, completely new hydrogen interaction potentials may be
relevant at high occupancies.  In any case, the high abundance of H
and/or H$_2$ can only be altered from that predicted previously if rapid
migration to the surface of H or newly formed H$_2$ becomes possible.
In fact, in refs. \cite{MS} it has been argued that there is strong
indication of H$_2$ formation and migration in some forms of graphite
and amorphous carbon films saturated with hydrogen.  There is
insufficient data to make a quantitative estimate of the rate for the
latter process.  It is likely, though, that once the bulk becomes
saturated, then a newly intercepted atom in the bulk will either
diffuse quickly to the surface where it evaporates or undergoes
quantum pair recombination, as discussed above, or reside in the bulk
long enough to recombine with another incident atom to form an H$_2$
molecule that will subsequently be ejected from the grain.  The rates
for these processes depend on the barriers of the shallow sites in the
saturated bulk and the modified activation energy for bulk to surface
transition.  In all these more complicated scenarios, H$_2$ formation
increases compared to the simple estimates based solely on pair
evaporation. Thus, we conclude that at low enough grain temperatures
effective H$_2$ formation should take place in or on these putative
carbon grains.

\section{Summary and Conclusions} \label{sec:sum}

In this paper we have examined the conditions required for effective
molecular hydrogen formation in and on solids, with particular
emphasis on astrophysical grains.

Much of the earlier work has been concerned with the catalysis of
H$_2$ formation by surfaces of grains in molecular clouds, where the
grain and gas temperatures are very low.  The key processes
determining the rate of H$_2$ formation are then: sticking and
retention of the gas-phase atoms, diffusion of H along the surface,
recombination with another H atom, and ejection of the newly formed
H$_2$ molecule.  In these scenarios the chemisorption sites are taken
to be saturated and hydrogen atoms and molecules are weakly bound to
physisorption sites with binding energies on the order of a few times
$10^{-2}$ eV \cite{HS}.  This path may apply to grains coated by overlayers of
molecular material such as amorphous ice \cite{BZ}.  
The evaporation time of H from these weak binding sites is
very short even at the low grain temperatures envisioned, but deeper
(impurity) sites may retain a fraction of the H atoms for long enough
time, and act as reaction sites \cite{HS}.  The migration time of a newly
adsorbed H atom to a reaction site is typically very short compared
with the residence time \cite{HS2} (but c.f.
ref. \cite{BZ}), so that the factors limiting the H$_2$ formation
rate are the sticking probability and retention of H.

The mechanisms controlling the rate of molecular hydrogen formation on
grains could be markedly different under harsher conditions, e.g., in
the vicinity of an astrophysical shock.  Firstly, the grain may be
exposed to fluxes of energetic particles or intense UV/X radiation,
and this may lead to surface cleaning, surface reshaping, lattice
modification and so forth that create chemisorption sites.  Secondly,
the gas-phase atoms may be energetic enough to penetrate deep inside
the grain and, under certain conditions, to drive direct collisional
reactions.  H$_2$ formation may then occur, (1) in the bulk via
thermal activation, (2) in the bulk by direct recombination with
incident H atoms, (3) on the surface by thermally activated
recombination of chemisorbed H atoms, (4) by direct recombination of
an incoming atom with a chemisorbed atom, and (5) through pair
evaporation.  In case 4 a fraction of the newly formed H$_2$ molecules
may be ejected immediately from the grain, a possibility discussed
earlier in ref. \cite{Du}, and referred to here, following him, as
prompt reaction. Chemisorption of hydrogen on carbon has been studied
by several authors \cite{Ar,Kl}, who
calculated interaction potentials and chemisorption energies of atomic
hydrogen on ideal graphite surfaces using quantum chemical
techniques. There remains some variation in these results, so we have
treated the value of the binding energy as a parameter.

The efficiency of H$_2$ formation depends, quite generally, on the
characteristics of the solid including the following energy scales:
the ground state energies of H and H$_2$ in the bulk, surface and
vacuum; the energy barriers for site to site diffusion, for surface to
bulk exchange, for H$_2$ formation in the bulk, on the surface and for
pair formation above the surface.  The efficiency also depends on the
grain temperature and the flux and energy (particularly if high enough
to drive collisional reactions) of H nuclei impinging on the grain. It
is important to remember that the energy levels and activation
barriers themselves depend on the density of hydrogen in and on the
solid by virtue of the self interaction of hydrogen atoms and
molecules. These values could be altered substantially when the grain
reaches saturation.

In this work we have explored various recombination reaction pathways
that may operate under a wide range of conditions.  By constructing
simple analytic models, we identify the regimes in the space of
physical parameters that correspond to {\it effective H$_2$ formation}
and {\it grain saturation} by H atoms and H$_2$ molecules.  Although
our model is general we have focused on the case of ``clean'' grains
with only one type of surface site and one type of bulk site.  So,
our examples do not apply to the classic molecule formation
mechanism explored by Hollenbach and Salpeter \cite{HS2} which envisioned
two surface sites with different binding energies.  
We also assume that hydrogen diffusion in the
bulk and on the surface is rapid enough to render the bulk and surface
occupation probabilities homogeneous.  The processes incorporated in
our model include: thermally activated and collisionally induced
exchange of H and H$_2$ between the bulk and the surface, H$_2$
formation in the bulk and on the surface by thermal activation and
direct recombination with incoming atoms, pair evaporation, prompt
reaction, thermal evaporation and collisional ejections and a variety
of dissociation reactions in and on the solid (by thermal activation,
by quantum tunneling and by direct collisions).

Despite the large number of physical parameters involved, the
solutions for the H and H$_2$ bulk and surface concentrations and the
H$_2$ formation efficiency are found to be highly degenerate, and
depend only on a relatively small number of dimensionless parameters.
In the following, we briefly summarize the main results for the three
distinct cases we have considered: when H$_2$ forms from H atoms leaving
the surface, when it also forms on the surface, and finally when it
forms in the bulk.

If there are no bound H$_2$ states in the bulk and on the surface,
H$_2$ formation occurs via pair evaporation or prompt reaction. Two
cases are of interest. (a) In materials which have positive
chemisorption energy, as appears to be the case for carbon, pair
evaporation can proceed quantum mechanically even at zero temperature.
Our result is that ${\cal E} > 0.5$ when $S ( (\Gamma_{2H}/\fracsurf
\Gamma_H^2) + 2 \alpha_{Fs}/\Gamma_H) > 1$. In the absence of
collisions this is simply $-\ln(\Gamma_{2H}/\nu_o) - 2E_e/kT_g - 
\ln( S/\fracsurf \nu_0) < 0$
so that effective H$_2$ formation is ``guaranteed'' as long as the
loss rate of atomic hydrogen by evaporation is less than the quantum
pair evaporation rate. This is essentially a condition on the maximum
grain temperature. (The threshold grain temperature does depend on the
rate at which H nuclei are intercepted by the grain but
the dependence is logarithmic and therefore weak.) If the grain
temperature is large, the surface density of H is small and the
efficiency of formation of H$_2$, which is a process quadratic in the
surface density, is small.  (b) For materials with negative
chemisorption energy, binding to the surface of H is energetically
favorable so that surface densities would saturate at typical grain
temperatures. Pair evaporation and H evaporation both require thermal
activation but generally molecule formation will proceed with more
rapidity simply because H$_2$ is bound with respect to H in the
vacuum. (We are assuming that the barrier to H$_2$ formation is less
than the embedding energy of H.)  For these materials, the efficiency
of molecule formation will be high. However, since the overall loss
rate is exponentially sensitive to temperature these materials will
become saturated if the grain temperature is too low. Saturation of
the surface occurs when $2\Gamma_{2H}+\Gamma_H<\fracsurf^{-1}S$.
Whenever grain saturation of the surface occurs it is possible that
the important parameters describing the grain are modified.

For both materials above, collisional ejections of H atoms and prompt
reaction may alter the results; the former releases atomic hydrogen
from the grain and reduces the H$_2$ formation efficiency, whereas the
latter leads to effectively higher surface recombination rate and,
therefore, enhanced efficiency. The rates of both processes are
independent of the grain temperature (to the extent that $\alpha_{Fs}$
and $\alpha_{sv}$ themselves do not depend on T$_{gr}$). As figure~\ref{fig:3}
illustrates, these processes can significantly alter the regimes of
effective and ineffective H$_2$ formation and also grain saturation.

When H$_2$ can associate/dissociate and adhere to the grain surface
(but not the interior) the above situation is more complicated.  H$_2$
formation on the surface always increases the formation rate by pair
evaporation alone [this is mathematically made clear by the simple
modification needed for eqn. (\ref{eq:NH sat1})].  The efficiency of
molecule formation is increased.  This qualitatively modifies our
previous example only when formation would be otherwise inefficient.
Efficient formation requires
$\bar{t}_{Fs}^{-1}>>\fracsurf\Gamma_H^2/S$, where $\bar{t}_{Fs}^{-1}$
is the effective formation rate defined explicitly in eq. (\ref{eq:a
coeff}). The limiting cases were discussed and correspond to (1) H$_2$
formation followed by immediate evaporation from the surface and (2)
evaporation of the surface equilibrium H$_2$ fraction. Numerically,
efficiency contours are given by the modified form of eq. (\ref{eq:eff
cont1}).

In so far as saturation is concerned, it is more difficult for H to
saturate the surface. At the same time, surface formation also
introduces the possibility that H$_2$ may become saturated on the
surface if the release rate from the grain is not fast enough. The
detailed conditions are given in eqn. (\ref{eq:sat H H2}).

H$_2$ production in the bulk requires the existence of bound H$_2$
states inside the solid, as depicted in figure~\ref{fig:2}. It is unknown which
substances allow such bound states. If bound states exist, the process
has many similarities to H$_2$ formation on the surface. Likewise,
it increases the net efficiency. We have calculated the effect of bulk
formation independent of the surface and pair evaporation processes.
The H and H$_2$ occupation probabilities as well as the formation
efficiency depend on the net rate of release of atomic and molecular
hydrogen from the bulk and, consequently, involve the exchange rates
between bulk and surface.  The full solution for H and H$_2$
occupancies and the efficiency depends on 7 dimensionless parameters
but the bulk concentration depends on only 2.  Additional
simplifications are possible if the transfer time of H from surface to
bulk is very short or the injection rate of H into bulk sites exceeds
that into surface sites. The latter case is particularly relevant for
homogeneous injection, the limit achieved when high energy particles
penetrate the lattice. We show that the bulk concentration of H may be
simply expressed as a function of $x_R$ and $x_F$ (figure~\ref{fig:5}), where $x_F$
is the ratio of bulk H$_2$ formation rate to injection rate per site,
and $x_R$ is the ratio of net release rate of H from the bulk to
injection rate per site. The sensitivity of these results to the
dissociation rate is also discussed (the ratio of H$_2$ dissociation
rate and release rate of H$_2$ from the bulk is $\eta$). The domains of
H and H$_2$ saturation are described.

H$_2$ formation within the bulk can proceed as a thermally activated
reaction or by collisional impact of a high energy particle.  If
collisional H$_2$ formation is less important than thermally activated
formation, then equilibrium is achieved between release and thermal
formation. The efficiency depends on $x_F/x_R^2$ (see figure~\ref{fig:6} and
segment with slope 2). In the alternative limit, collisional formation
achieves equilibrium with release (figure~\ref{fig:6} with segment with slope 0).
The location of the efficiency curves also depends on $\eta$.  The
efficiency is essentially independent of $\eta$ when $\eta<<1$ and
decreases as $\eta$ increases for $\eta>1$.

As an application we have considered H$_2$ formation on carbon.  This
system is complicated by virtue of the broad spectrum of binding sites
exists in the solid. Despite this complication we were able to draw
some general conclusions using results of simulations and data 
available in the literature.  In particular,
1) efficient molecule formation requires grain temperatures
smaller than about 100 K, 2) pair evaporation will keep the surface
unsaturated even at very low grain temperatures if 
$n$(H)($\epsilon/\eV$)$^{1/2}<2.5\times10^9$ cm$^{-3}$ where $n$(H) is
the number density of gas phase atoms and $\epsilon$ is their average 
energy, and 3) the bulk will be saturated in steady state
unless sufficiently rapid molecule formation will take place in 
the bulk followed by rapid transfer of H$_2$ to the surface or  
if the incident atoms are energetic enough to eject bulk H to the
surface at a large enough rate.                                             

\acknowledgements

D.F.~Chernoff thanks Chris McKee for stimulating conversations which
motivated this work. We acknowledge helpful discussions with Neil
Ashcroft, Barbara Cooper, Tatjana \'{C}ur\v{c}i\'{c}, Bruce Roberts
and Mike Teter.  This work was supported by an Alon fellowship (AL)
and at Cornell University by NSF grant AST-9530397 and NASA grant
NAGW-5-2851.

\newpage

{\bf FIGURE CAPTIONS}
\begin{figure}
\caption{Schematic energy diagrams of hydrogen in solids 
for which no bound H$_2$ states exist in the bulk and on the surface.  
The solid curve 
represents the interaction potential of a single H atom with
the solid.  The two dashed lines represent the interaction 
potential of an atom pair; one dashed line corresponds to a pair 
at fixed interatomic distance of order that of a free H$_2$ molecule, 
and the other to a minimum energy configuration.  The chemisorption 
energy, $E_c$, solution energy, $E_s$, activation energy of 
chemisorption, $E_a$, activation energy of pair evaporation, $E_r$, and
dissociation energy of H$_2$ in vacuum per H atom are indicated.  The 
sign of $E_c$ and $E_s$ is defined by the direction of the 
corresponding arrows; positive if the arrow points upwards and negative
if it points downwards.  The four cases shown correspond to different 
possible ordering of the energy levels (see text for detailed discussions).}
\label{fig:1}
\end{figure}

\begin{figure}
\caption{The same as figure~\ref{fig:1}, but in the case in which H$_2$ 
is stable in and on the solid.  The thick arrows indicate different 
pathways for H$_2$ formation and release: [1] $H_2$ formation in the bulk,
[2] H$_2$ formation on the surface, [3] pair evaporation, and [4] 
evaporation of H$_2$ from the surface.  Each reaction may involve 
an activation barrier as shown schematically.}
\label{fig:2}
\end{figure}

\begin{figure}
\caption{Domains of grain saturation, effective and 
ineffective H$_2$ formation for materials for which there are 
no bound H$_2$ states on the surface and in the bulk, in the absence of
(windows a and b) and in the presence of (windows c, d) collisional 
reactions.  The independent variables are the log of pair evaporation 
rate per attempt frequency, $\Gamma_{2H}/\nu_o$, and the embedding energy 
in units of the grain temperature $E_e/kT_g$. } 
\label{fig:3}
\end{figure}

\begin{figure}
\caption{Conditions for effective and ineffective H$_2$ 
formation for specific materials.  The dashed line corresponds 
to embedding energy $E_e= 2.55$ eV, chemisorption energy $E_c=-0.3$ eV, 
activation energy of chemisorption $E_a=0.2$ eV, and barrier 
penetration depth $G_H>7$, and represents a typical transition 
metal.  The dotted-dashed line corresponds to $E_c>0$, $E_e=0.3$ eV,
activation energy of pair evaporation $E_r=0.4$ eV, and $G_H=30$, 
and represents carbon (see example in \S \ref{sec:app}). }
\label{fig:4}
\end{figure}

\begin{figure}
\caption{$\nbulk_H$ contours plotted in the plane defined 
by $x_F$ - the ratio of thermally activated H$_2$ formation
rate in the bulk and injection rate per site, and $x_R$ - the 
net release rate per H atom of bulk atomic hydrogen in units 
of the injection rate per site, for $\alpha_{Fb}=0$, $\eta=0$ 
(solid lines) and $\eta=10^2$ (dotted lines).  Values of $\nbulk_H$ 
label the curves.  The curve separating the regimes of saturated and 
unsaturated grain is shown schematically (dashed line).}  
\label{fig:5}
\end{figure}

\begin{figure}
\caption{Contours of H$_2$ formation efficiency, 
${\cal E}$, for $\eta=0$, $\alpha_{Fb}=0$ (solid lines)
and $\alpha_{Fb}=10^2$ (dashed lines).  The efficiency 
curves are broken power laws (see eq. [\ref{eq:cont eff b}]).  
The change in slope reflects the transition from the regime where
thermally activated H$_2$ formation dominates to the regime where 
collisional recombination of a bulk H atom with an incoming 
atom dominates. }
\label{fig:6}
\end{figure}

\newpage
{\bf APPENDIX: LIST OF SYMBOLS}

\noindent $T_g$ - Grain temperature

\noindent $T_t$ - Tunneling temperature

\noindent $\nbulk_{X}$ - Fraction of all sites occupied by species $X$ in the bulk; ($X=$H, H$_2$)

\noindent $\nsurf_{X}$ - Fraction of all sites occupied by species $X$ on the surface; ($X=$H, H$_2$)

\noindent $\ntotal$ - Total number of sites 

\noindent $\fracbulk$ - Ratio of the number of bulk sites and $\ntotal$

\noindent $\fracsurf$ - Ratio of the number of surface sites and $\ntotal$

\noindent $S$ - Total injection rate per site

\noindent $S_H$ - Injection rate per site into bulk sites

\noindent ${\cal S}_H$ - Injection rate per site into surface sites

\noindent ${\cal E}$ - H$_2$ formation efficiency

\noindent $E[X]^i$ - Ground state energy of species $X$; ($X=$H, H$_2$, $i=b, s, v$)

\noindent $E_c$ - Chemisorption energy

\noindent $E_e$ - Embedding energy

\noindent $E_s$ - Solution energy 

\noindent $E_d^i$ - Dissociation energy; ($i=b, s, v$)

\noindent $E_r$ - Activation energy of pair evaporation 

\noindent $E_a$ - Activation energy of chemisorption

\noindent $G_H$ - Penetration factor for quantum pair evaporation

\noindent $t_{in}$ - Mean time for particle interception by the grain

\noindent $t_{Fi}$ - Thermally activated H$_2$ formation time; ($i=b, s$)

\noindent $t_{Di}$ - Thermally activated H$_2$ dissociation time; ($i=b, s$)

\noindent $t_{Fsb}$ ($t_{Fbs}$) - Time of H$_2$ formation in the bulk (on the surface) 
by thermally activated recombination of surface and subsurface atoms

\noindent $t_{Dbs}$ ($t_{Dsb}$) - Time of H$_2$ dissociation in the 
bulk (on the surface) to produce a surface and subsurface H

\noindent $t_{sbX}$ ($t_{bsX}$) - Time of thermally activated diffusion of a species $X$ 
from the surface (bulk) to the bulk (surface)

\noindent $\alpha_{Fi}$ - Probability per site of direct collisional recombination with an incoming H 
atom; ($i=b, s$)

\noindent $\alpha_{svX}$ - Probability per site of collisional ejection of a species $X$

\noindent $\beta_i$ -  Probability per site of collisional dissociation of an H$_2$ molecule; ($i=b, s$)

\noindent $\gamma_{X}$ - Thermal evaporation rate of a species $X$ from the surface

\noindent $\Gamma_{X}$ - Total loss rate of a species $X$ from the surface (sum of thermal evaporation 
and collisional ejection)

\noindent $\Gamma_{2H}$ - Pair evaporation rate

\noindent $q$ - Fraction of collisional dissociations on the surface that results in the immediate loss of
the H atoms

\noindent $\delta_{Fs}$ - Fraction of H$_2$ molecules formed by collisional recombination on the surface
that remain on the surface

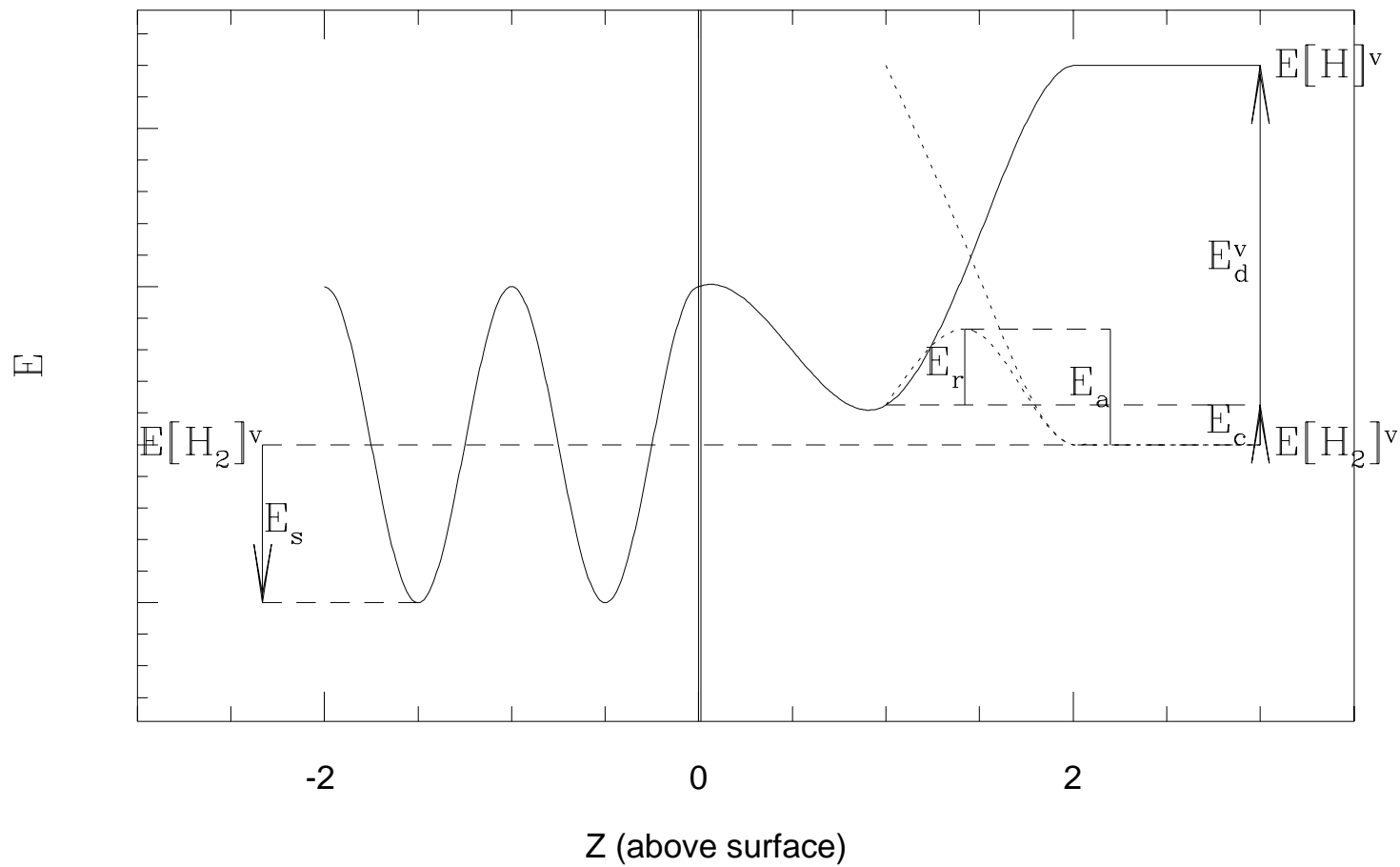
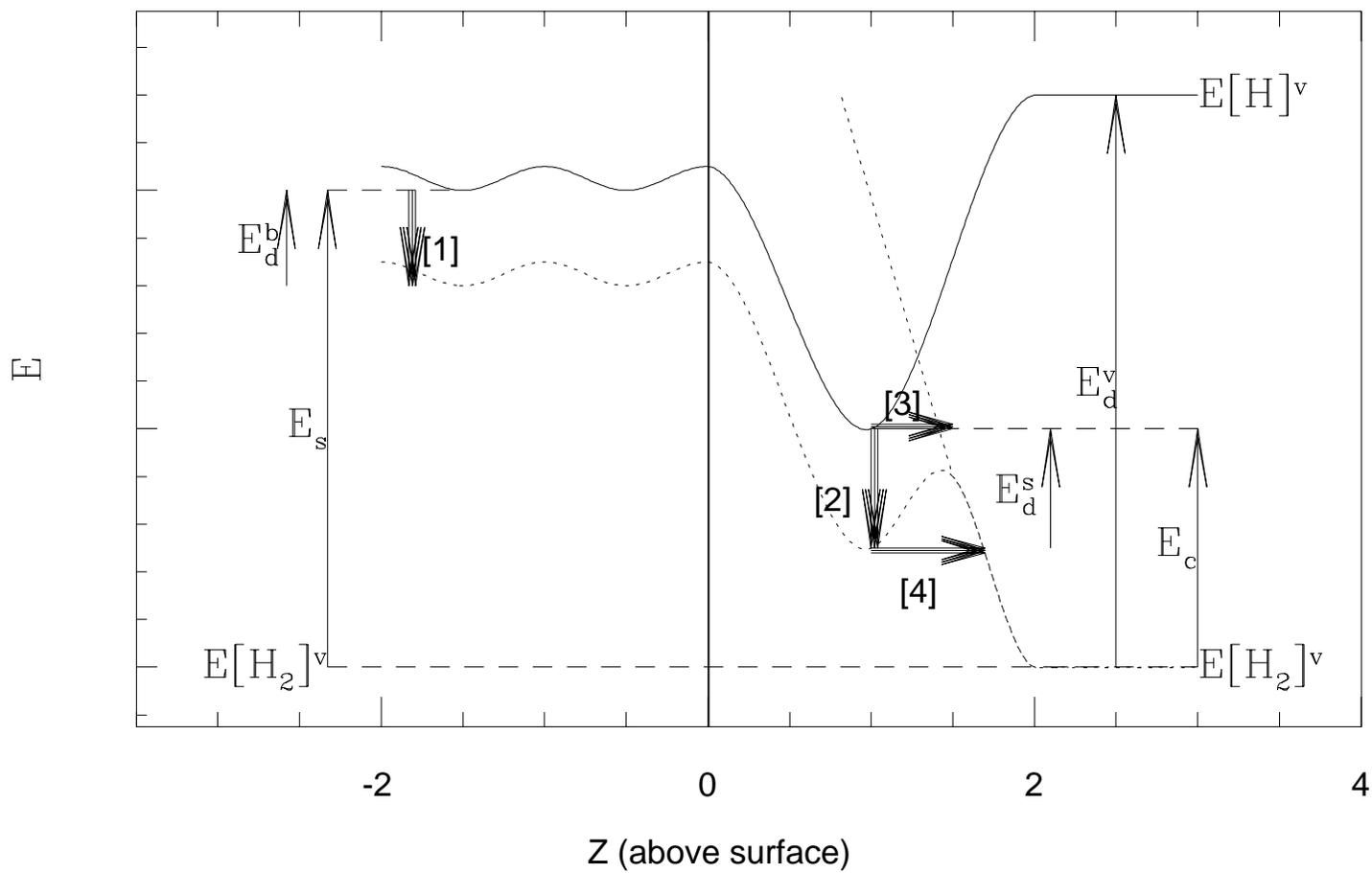
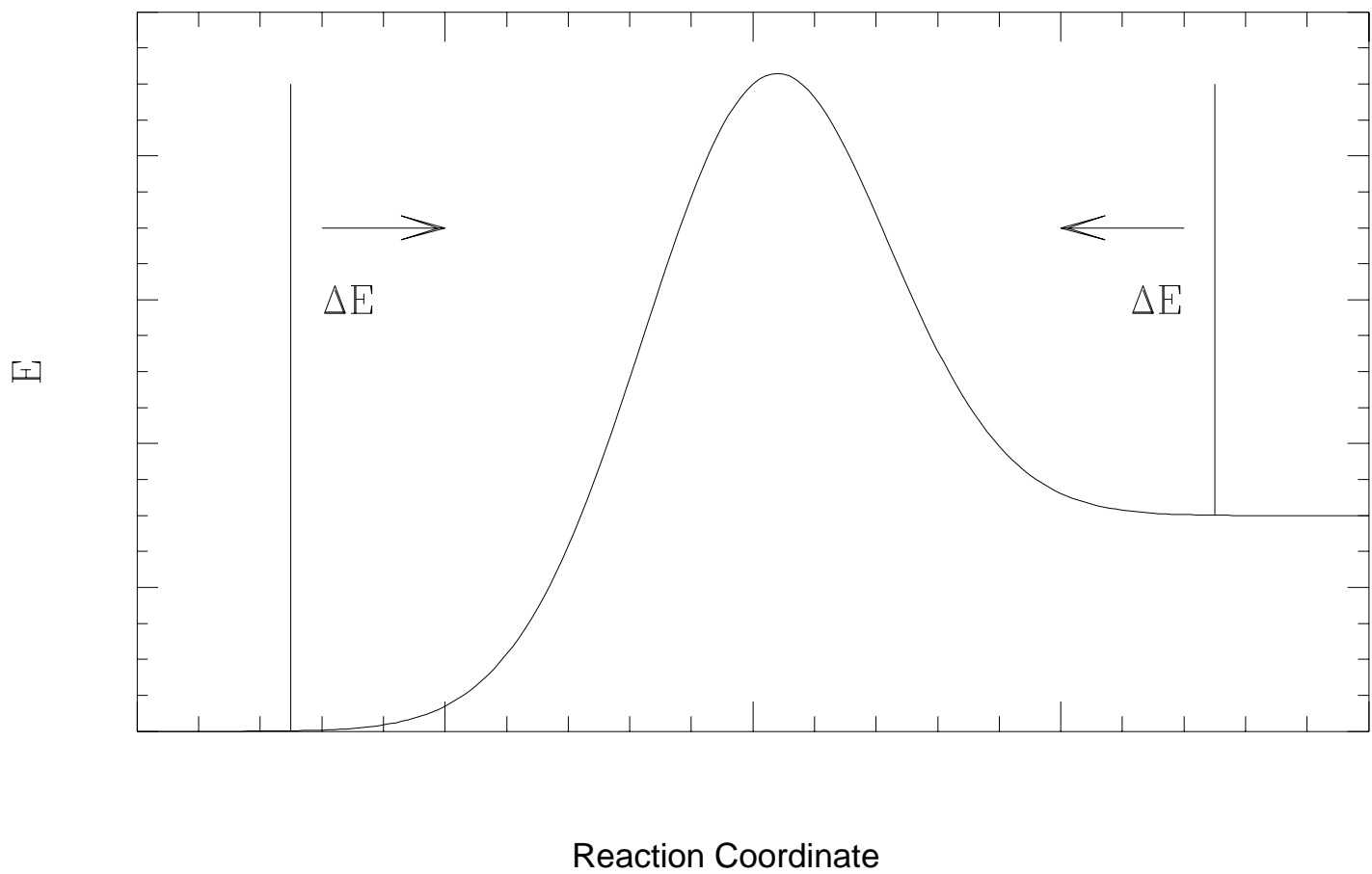
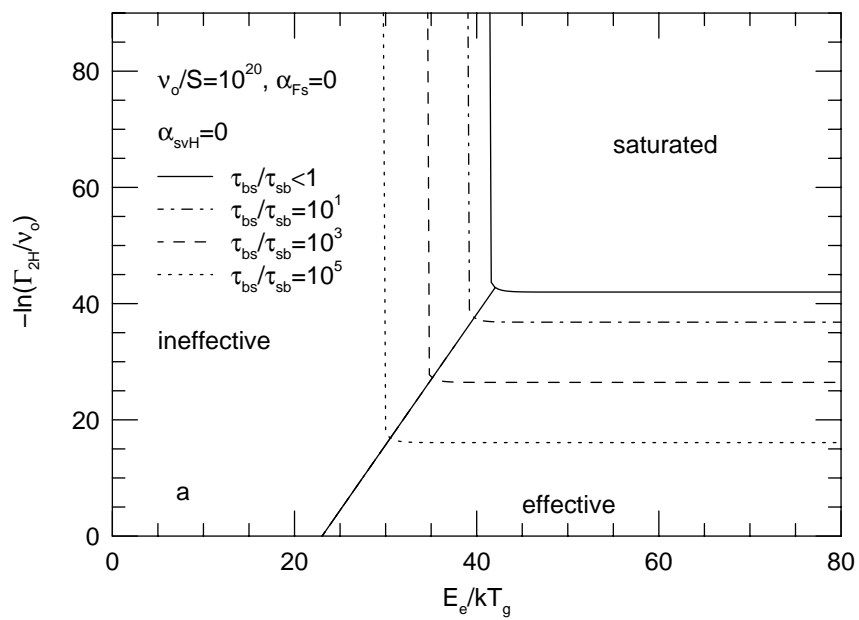
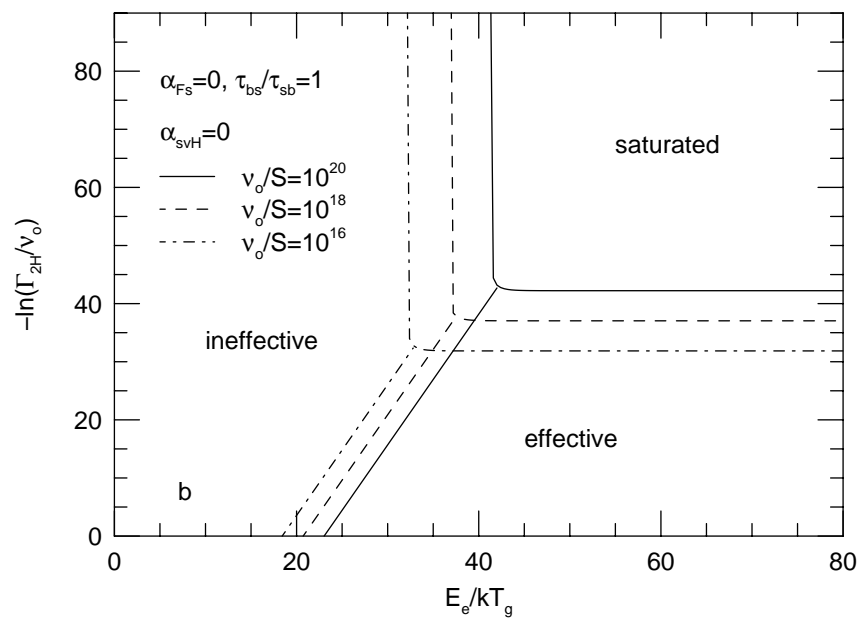
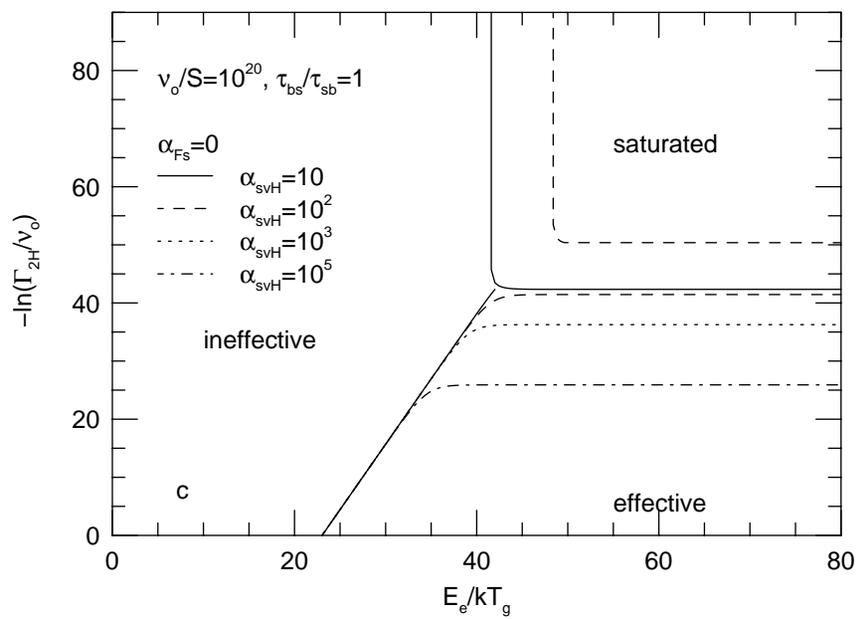
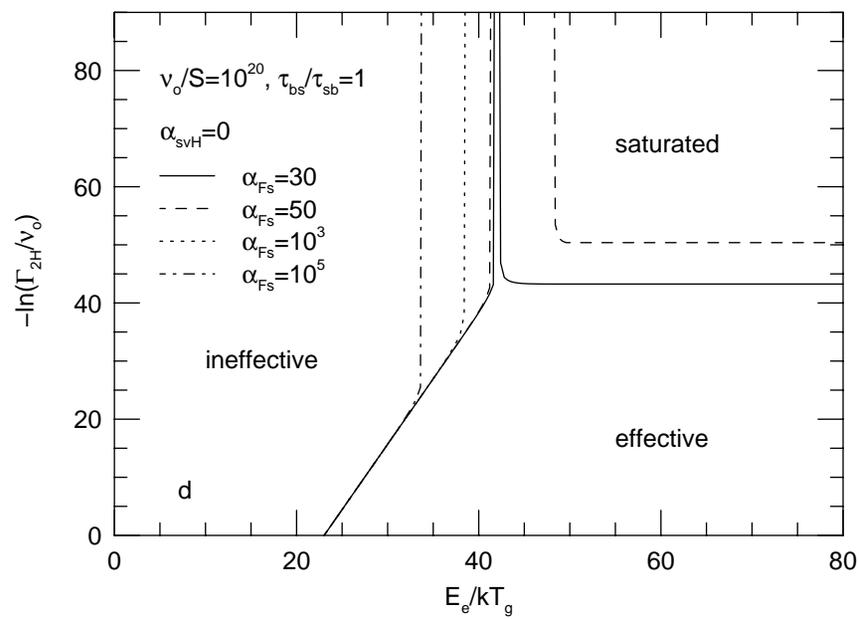
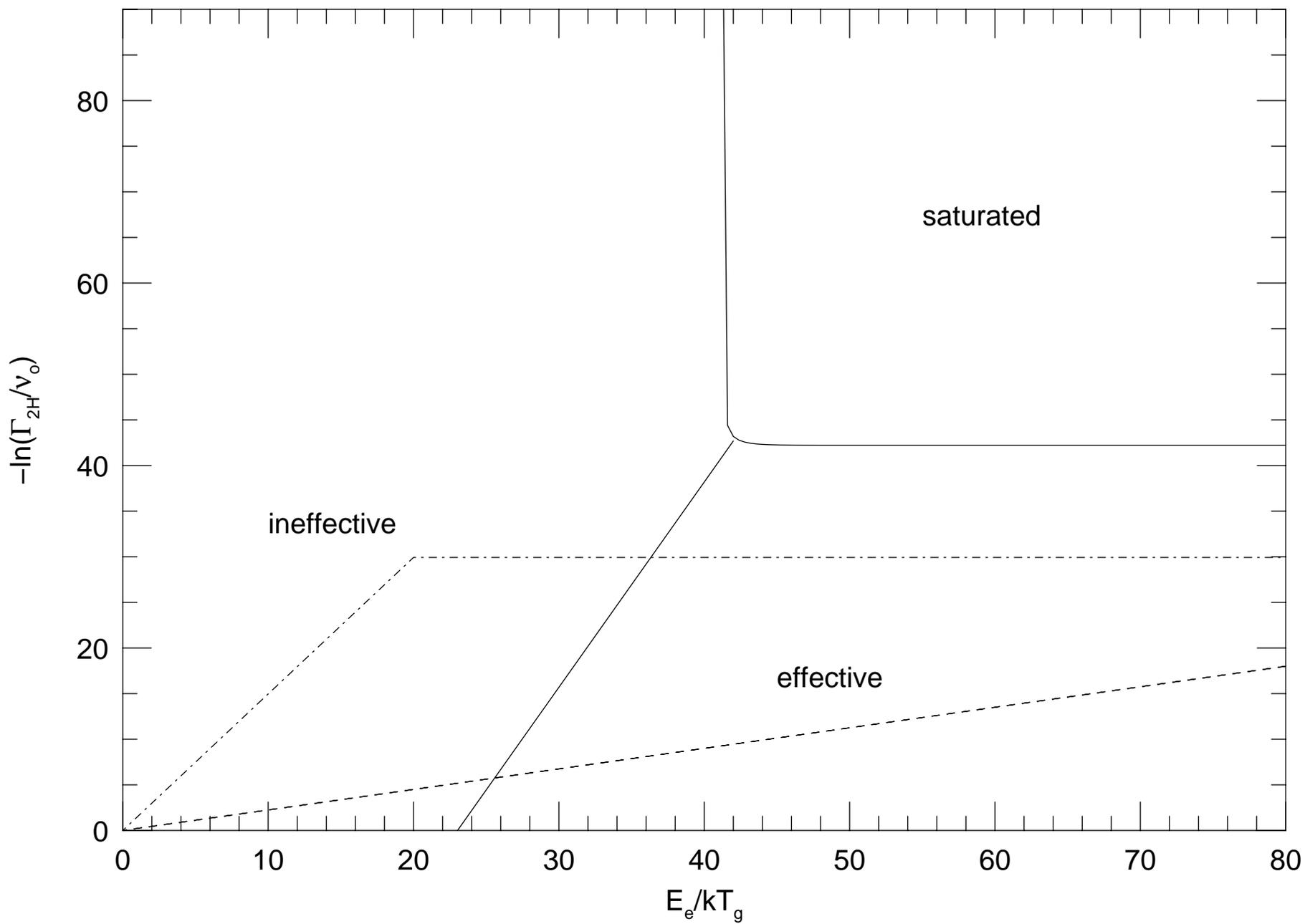
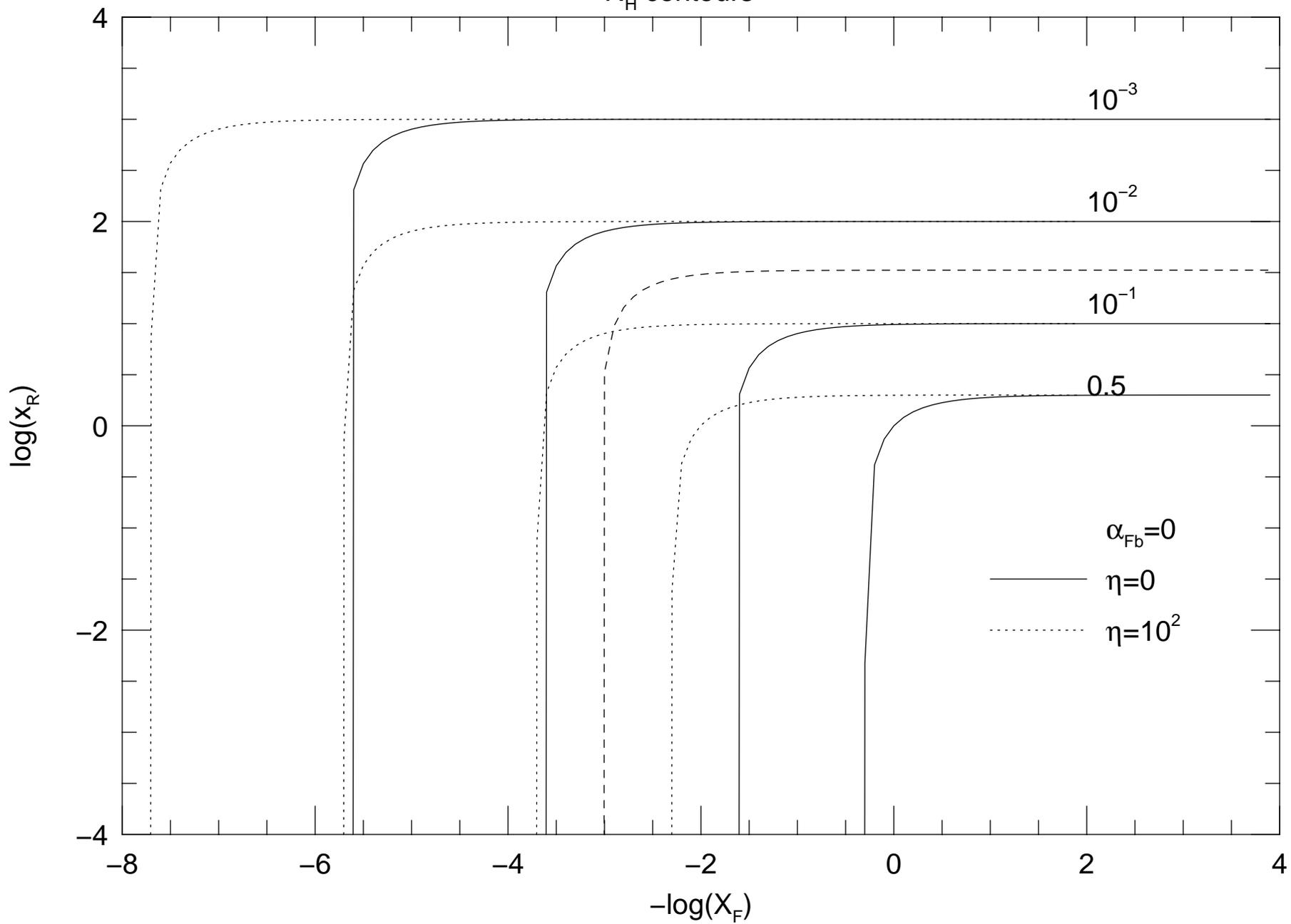
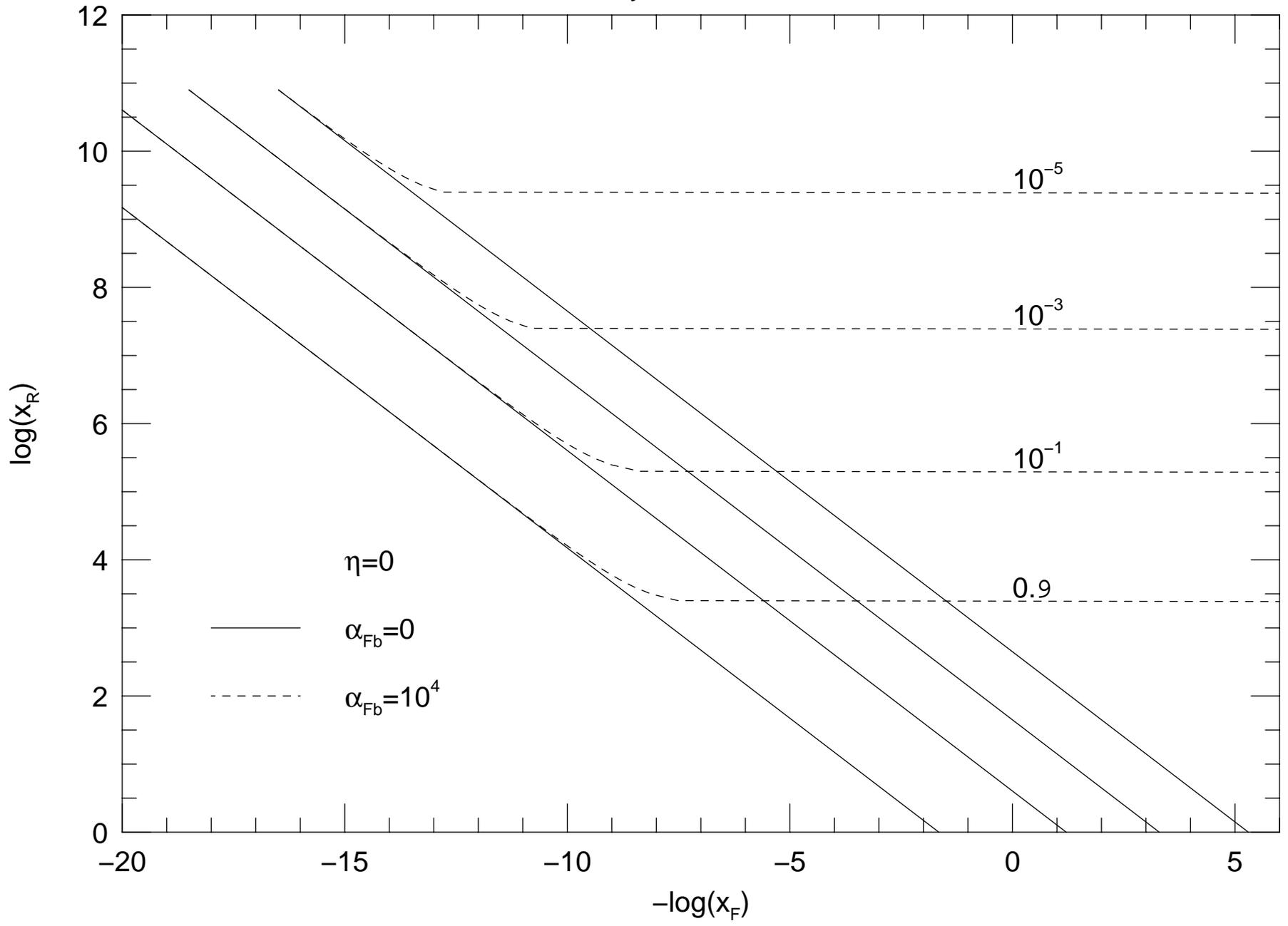
\end{document}